\documentclass[aps,prb,twocolumn,preprintnumbers,amsmath,amssymb, groupedaddress]{revtex4}
\usepackage{dcolumn}% Align table columns on decimal point
\usepackage{bm}% bold math
\usepackage{booktabs} %tables
\usepackage{graphicx}% Include figure files

\begin{document}

\title{Existence of a Phase with Finite Localization Length in the Double Scaling Limit of N-Orbital Models}

\author{Vincent E. Sacksteder IV}
\email{vincent@sacksteder.com}
\affiliation{Department of Physics, Royal Holloway University of London, Egham Hill, Egham, TW20 0EX, United Kingdom}
\affiliation{Department of Physics and Astronomy, Rutgers University, NJ, USA}

%\author{Tomi Ohtsuki}
 %\affiliation{Sophia University }

\date{\today}% It is always \today, today,

\begin{abstract}
%  We treat models of disordered conduction with $N$ orbitals per site, which are important  for their improved mathematical tractability compared to other disordered models, their close connection to sigma models, and their physical application to systems of coupled disordered grains.
Among the models of disordered conduction and localization, models with  $N$ orbitals per site are  attractive both for their mathematical tractability and for their physical realization in coupled disordered grains.  However Wegner proved that there is no Anderson transition and no localized phase  in the $N \rightarrow \infty$ limit, if the hopping constant $K$ is kept fixed. \cite{PhysRevB.19.783,Khorunzhy92} Here we show that  the localized phase is preserved in a different limit where $N$ is taken to infinity and the hopping $K$ is simultaneously adjusted to keep $N \, K$ constant.   We support this conclusion with two arguments.  The first is  numerical computations of the localization length showing that in the $N \rightarrow \infty$ limit the site-diagonal-disorder model  possesses a localized phase if $N\,K$ is kept constant, but does not possess that phase if $K$ is fixed.  The second argument is a detailed analysis of the energy and length scales in a functional integral representation of the gauge invariant model.  The analysis shows that in the $K$ fixed limit the functional integral's  spins do not exhibit long distance fluctuations, i.e. such fluctuations are massive and therefore decay exponentially, which signals conduction.  In contrast the $N\,K$ fixed limit preserves  the massless character of certain spin fluctuations, allowing them to fluctuate over long distance scales and cause Anderson localization.
\end{abstract}

%\pacs{?, 72.15.Rn, 64.60.De, 72.20.-i, 64.60.Cn}
% submit to Journal of Physics A: Mathematical and Theoretical

\maketitle

\section{The Site-Diagonal-Disorder Model}

 In the first sections of  this article we study Wegner's site-diagonal-disorder  model with $N$ orbitals on each site, which at $N=1$ reduces to the Anderson model. \cite{PhysRevB.19.783,Anderson58}  The on-site Hamiltonian $H_{i}$ for site $i$ is an $N \times N$ random matrix which describes interactions between the $N$ orbitals at  site $i$.  It is  the natural $N \times N$ generalization of the Anderson model's disorder potential.     The mean squared value of $H_{i}$'s matrix elements is proportional to $N^{-1}$, so that when a single site is taken in isolation, its spectral density (except for tail states) is in the interval $E = \left[-2,+2\right]$.  This interval is independent of $N$'s value.   The   averaged eigenvalue density $\nu(E)$ of a single site obeys the semicircle law,
\begin{equation}
\label{eq:defRandomMatrix}
\nu(E)=N \;\frac{\sqrt{4-E^2}}{2\pi}\,.
\end{equation}

Like the Anderson model, the site-diagonal-disorder model combines random on-site disorder with deterministic hopping between sites.  The site-diagonal-disorder model's  hopping  connects each site with its nearest neighbors.  The hopping matrix is  proportional to  the $N\times N$ unit matrix $I_N$, i.e.
\begin{equation}
\label{eq:siteHopping}
H_{i,j}=K\; I_N\,,
\end{equation}
where $K$ is the hopping strength, and $i,j$ are the position indices of two nearest neighbors.   This hopping matrix is a natural $N \times N$ generalization of the Anderson model's non-random hopping.

 We will examine the site-diagonal-disorder model with  two alternative symmetry classes. The first is the orthogonal  symmetry class with real matrices obeying $H_{i} = H_{i}^T$, and the second is the unitary symmetry class with  Hermitian matrices $H_{i} = H_{i}^\dagger$.  The orthogonal class describes systems with time reversal symmetry, while the unitary class describes systems where time reversal symmetry is broken, for instance by a magnetic field. 

Our numerical work will concentrate on the one dimensional case where the sites are arranged in a long one dimensional chain, and the position of each site is denoted by the index $i$.    The nearest neighbors of the $i$-th site are at $i\pm1$.

 Working under the implicit assumption that the hopping strength $K$ between sites is kept fixed as $N$ is varied, Wegner showed that in the $N \rightarrow \infty$ limit the site-diagonal-disorder model always has finite conductivity; it is always in the conducting phase, for all energies within the band. \cite{PhysRevB.19.783} Later Khorunzhy and Pastur supplied a more mathematically precise proof. \cite{Khorunzhy92} 
This result implies that the localization length $\xi$ is infinite in the $K$ fixed, $N \rightarrow \infty$ limit.  Our numerical results in section \ref{KFixedNumerics}  confirm the infinite localization length and absence of localization in the $K$ fixed, $N \rightarrow \infty$ limit.  

This result, the absence of a localized phase in an  disordered model with short-range hopping and no disorder correlations, is a bit peculiar because it is independent of the dimensionality.  In particular, Wegner's proof of the absence of a localized phase applies to 1-D wires.  His proof contrasts with rigorous analytical proofs that 1-D wires with uncorrelated disorder and short range hopping are localized,  \cite{ishii1973localization,kramer1993localization}
% See also, for continuous problems, Golds’heidIYa, Molchanov S A and Pastur L A 1977 Funkt. Anal. Prilozken 111, and also  Molchanov S A 1978 Math USSR Izoestija 42 trad 1269
except in the special case of certain ensembles with perfectly conducting channels.  \cite{PhysRevLett.69.1584,Mirlin94,Ando02,Takane04,PhysRevLett.99.116601,PhysRevLett.99.036601,ashitani2012perfectly}  Included among such proofs is a recent one concerning  the same site-diagonal-disorder model discussed here, in 1-D wires,  and proving that the model's spectrum is pure point if the number of orbitals $N$ is kept fixed. \cite{brodie2020eigenvalue}

The object of this paper is to show that Wegner's absence of a localized phase is a peculiarity of the limit that Wegner took, fixing the hopping strength $K$ while taking $N$ to infinity.  In section \ref{NKFixedNumerics} we will show numerically that in the double scaling limit where the product of $N$ and $K$ is kept fixed while taking  $N$ to infinity,  the site-diagonal-disorder model does have a  localized phase where the localization length is finite.  In other words, the $K$ fixed, $N \rightarrow \infty$ limit deletes the physics responsible for Anderson localization, while the $N\,K$ fixed, $N \rightarrow \infty$ double scaling limit preserves that physics.  

In subsequent sections we will analyze Wegner's gauge-invariant model, which differs from the site-diagonal-disorder model only in that the hopping between sites is random.  \cite{PhysRevB.19.783}After writing a functional integral which is mathematically equivalent to the gauge-invariant model, we will use  the saddle point approximation and perturbation theory to analyze the model's excitations and their energy and length scales.  Our analysis will show that in the $N \, K$ fixed limit certain excitations are massless and thus can cause localization, while in the $K$ fixed limit those excitations are made massive with a mass proportional to $N$.  Thus in the $K$ fixed limit these excitations are removed from the model, forcing it to conduct.

 A similar result about $N$-orbital models has recently been presented, proving exponential decay of matrix elements of the resolvent in a finite volume, which is a signature of Anderson localization. \cite{peled2019wegner}  Notably, the authors showed that exponential decay could be proved when $N \, K$ is smaller than a bounding function, pointing to the double scaling limit discussed here. Their result contrasts with ours in two respects: theirs concerns the decay of the resolvent while ours concerns the localization length, and theirs is a fully rigorous mathematical proof while we rely on both numerical calculations and a non-rigorous analysis of a functional integral which is rigorously equivalent to an $N$-orbital model.   Another difference is our focus directly on the model's behavior in the $N \rightarrow \infty$ limit where $N\,K$ is kept fixed; we perform this  limit numerically, and we use  $N\,K$ as a control parameter for perturbative analysis of the functional integral.
 
  \section{Numerical calculation of the localization length when the hopping strength $K$ is kept fixed.\label{KFixedNumerics}}
  
\begin{figure}%[tbhp]
\centering
\includegraphics[width=1\linewidth,clip,angle=0]{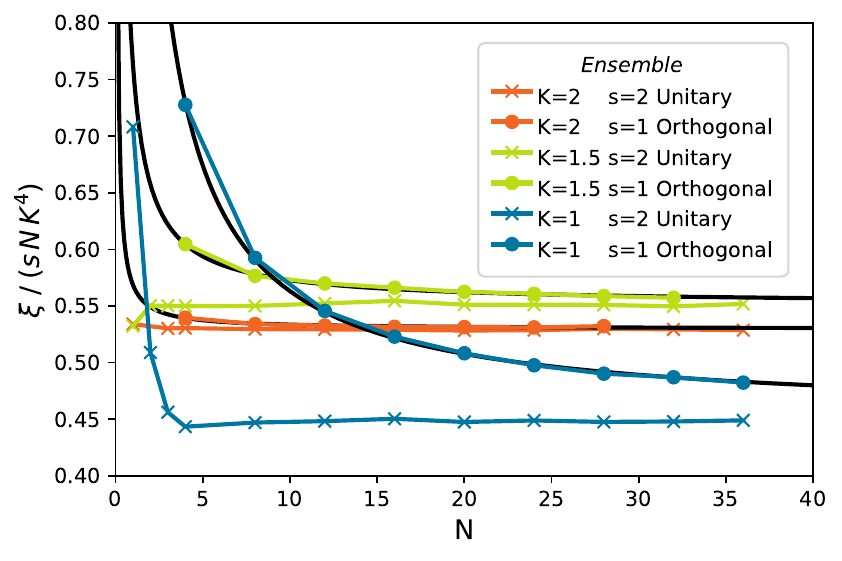}
\caption{ (Color online.)    The ratio of the localization length $\xi$ to the number of orbitals $N$, normalized by $s\,K^4$, in the $K$-fixed limit.   The ratio's convergence to fixed values at large $N$ shows that  in the  $K$-fixed  limit  the localization length $\xi$ is proportional to $N$. In consequence $\xi$ diverges and there is no insulating phase when $N$ is taken to infinity. Black lines are fits of $\xi$ to linear functions of $N$.
}
\label{figKFixed}
\end{figure}

    In Figure \ref{figKFixed} we present the localization length $\xi$, which we calculate using the transfer matrix method \cite{MacKinnon81}.  The transfer matrix method for calculating  localization lengths operates by building a  long chain with length $L$ and measuring the average decay within the chain.    We keep the  energy fixed at $E = 2.8$, fairly close to the band edge which is near $E = 3.5$ in the 1-D chain, in order to reduce the scattering length and the computational cost.  
    
%    shouldn't the band edge location depend on the disorder strength? the band edge which was around $3.5$ instead of $2$ because there is hopping.
    
We calculate $\xi$ with the hopping strength $K$ fixed at three values $K=1,2,3$, and for both the unitary and orthogonal ensembles.   In each of these six cases we calculate $\xi$ at several values of $N$ ranging up to $N=36$.  This upper limit on $N$ was determined by the numerical cost, which grows with $N$ for two reasons.  The first reason is that the numerical cost of adding one site to the chain scales with $N^3$.  The second is that, in the $K$-fixed limit, the scattering length and localization length both increase with $N$, so longer chains are required to obtain acceptable numerical accuracy.  We extended the chains until the statistical error of the localization length, quantified as a standard deviation, was reduced to 0.2\% percent of the localization length's value.
At  large $K=2.0$ and large $N=36$ this standard of numerical accuracy requires chain lengths $L$  longer  than  $10^8$ sites.

Figure \ref{figKFixed}  presents the ratio of the localization length $\xi$ to the number of orbitals $N$.  We normalize $\xi$ by the fourth power $K^4$ of  the hopping strength $K$, which is kept fixed at $1$, $2$, or $3$.  To compare the unitary ensemble to the orthogonal ensemble in the same graph, we divide the unitary results by $s=2$ and the orthogonal results by $s=1$.  

Each of the six data sets converges to a constant between $0.40$ and $0.60$ as $N$ becomes large, which shows that when $N$ is large the localization length $\xi$ grows linearly with $N$.   The black lines are fits of the orthogonal ensemble data to the linear relation $\xi(N) = a_0 + a_1 N$.  $a_1$ is the proportionality constant between $\xi$ and $N$, and $a_0$ is the extrapolated value of $\xi$ at $N=0$.    The numerical values of these fitting parameters are given in Tables \ref{tab:fitsParametersKFixedUnitary} and \ref{tab:fitsParametersKFixedOrthogonal}.  Figure \ref{figKFixed} shows excellent agreement between the linear fits and the data, and $\chi$-squared analysis also confirms that the fits are good.  
\footnote{It is interesting that the next-to-leading order term $a_0$ in the localization length, i.e. the first correction, is of order $O(1)$ in the orthogonal ensemble, but much smaller  $O(0.1)$ in the unitary ensemble. The unitary curves are basically flat for all $N \geq 4$, meaning that for these $N$ the $a_0$ term and any other corrections such as $1/N$ are very small. This may be a hint that in the perturbative expansion cancellations occur for the unitary ensemble but not for the orthogonal ensemble.}
This gives additional evidence that at large $N$ the localization length $\xi$ is proportional to $N$.

Since $\xi$ is proportional to $N$ at large $N$, when $N$ is taken to infinity the localization length diverges and there is no localized phase.  Since this numerical result was obtained in one-dimensional systems, and localization effects are generally strongest in 1-D and weaker in higher dimensions, it is a strong indication that the localized phase is absent in all dimensions.   This is  consistent with Wegner's demonstration, based on perturbation theory, that the site-diagonal-disorder model is always in the conducting phase in the limit of $N \rightarrow \infty$ with $K$ fixed.

 % Orth, listing the percent error and the chain length
  %K1  N=4  0.6% 0.7 mil, N=36 3.5% 6 mil
%  K15 N=4 2.5% 4 mil, N=32 18% 37 mil
%  K2  N=4 7%  17 mil, N=28 48% 131 mil	

% Unitary
%  K1  N=4  0.3% 0.2 mil, N=36 6.5% 11 mil
%  K15 N=4 1.1% 1.7 mil, N=36 40% 75 mil
%  K2  N=4 14%  35 mil, N=36 121% 323 mil	

%Orth
%  NK1 N=4 0.4% 0.4 mil, N=128 0.1%  0.06 mil
%  NK2 N=4 1% 1mil, N=128 0.2% 0.1 mil
%  NK3 N=4 2% 2mil, N=128 0.2% 0.2 mil
  
%  Unitary
%  NK1 N=4 0.5% 0.5 mil, N=128 0.1% 0.06 mil
%  NK2 N=4 1.6% 2mil, N=128 0.2% 0.1 mil
%  NK3 N=4 4% 4mil, N=128 0.2% 0.2 mil
 
\begin{table}
% These were taken from fittingParameters.xls, under localizationLength.
  \centering 
  \caption{Linear fitting parameters, $K$ fixed, Unitary Ensemble.  The data that was fitted had statistical errors of about 0.2\%.}\label{tab:fitsParametersKFixedUnitary}
   \begin{tabular}{|c|c|c|clcl}
\hline
  $K$  & $a_0$, Unitary & $a_1$, Unitary  \\ \hline
 1     & -0.0453032 & 0.899264 \\ \hline
1.5      & -0.0614256 & 5.58706  \\ \hline
2     & 0.212931    & 16.9203   \\ \hline
\end{tabular}
\end{table}

\begin{table}
% These were taken from fittingParameters.xls, under localizationLength.
  \centering 
  \caption{Linear fitting parameters, $K$ fixed, Orthogonal Ensemble}\label{tab:fitsParametersKFixedOrthogonal}
   \begin{tabular}{|c|c|c|clcl}
\hline
  $K$  & $a_0$, Orthogonal & $a_1$, Orthogonal  \\ \hline
 1     & 1.109 & 0.452275 \\ \hline
1.5      & 1.06834 & 2.7929  \\ \hline
2     & 0.622876    & 8.47447  \\ \hline
\end{tabular}
\end{table}

   \section{Numerical calculation of the localization length in the double scaling limit where $N K$ is kept fixed. \label{NKFixedNumerics}}

\begin{figure}%[tbhp]
\centering
\includegraphics[width=1\linewidth,clip,angle=0]{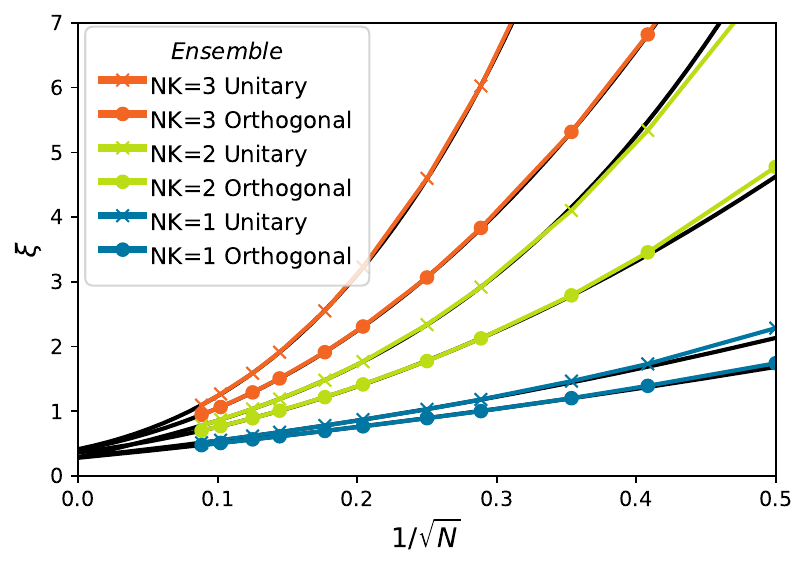}
\caption{ (Color online.)    Localization length $\xi$ as function of $N$, in the double scaling limit where $NK$ is held fixed.  The black lines are fits to quadratic or cubic polynomials in $1/\sqrt{N}$.  The points where the fits intercept the y-axis give the $N\rightarrow \infty $ limit of the localization length.
}
\label{figNKFixed}
\end{figure}

  Now we turn to the more interesting case where the product $N K$ is kept fixed while taking the $N \rightarrow \infty$ limit.  Figure   \ref{figNKFixed} shows the localization length $\xi$ with $NK$ fixed at $NK = 1, 2, 3$ for both the unitary and orthogonal ensembles.  When $NK$ is fixed to these values the scattering length  always remains  small and does not significantly affect the computational cost of calculating the localization length.  Therefore we fix the energy at the band center $E=0$ and calculate  $\xi$ at large values of $N$ up to $N=192$.

  Figure   \ref{figNKFixed} plots the numerical data on the localization length as colored symbols connected by line segments.    In the $NK$ fixed limit the localization length  $\xi$ does not diverge with $N$, and instead converges to a finite and non-zero value at $N \rightarrow \infty$. Therefore  Figure   \ref{figNKFixed}  plots $\xi$ vs. $N^{-1/2}$, so that the left border of the graph corresponds to $N\rightarrow \infty$.  This graph gives visual proof that $\xi$ does not diverge with $N \rightarrow \infty$.  If it did diverge, then $\xi$'s slope would increase as it approached the left border of the graph.  Moreover  $\xi$ itself would become either very large or very small near the left border and would escape the upper or lower boundary of the graph.   This is not the case:   $\xi$'s slope becomes smaller near the left border, and all curves approach  values in the interval between zero and one.

  Figure   \ref{figNKFixed} plots also black lines which  are fits of the data to  polynomial functions of $N^{-1/2}$. Both  $\chi$-squared analysis and visual examination show that these polynomials are excellent descriptions of  the localization length at small $1/\sqrt{N}$.  The fitting formula is 
  \begin{equation}
\label{eq:sqrtNFit}
\xi=b_0+b_1(NK)\; N^{-1/2}+b_2(NK) \;N^{-1}+b_3(NK) \;N^{-3/2}.
\end{equation}
The fitting parameters are listed in   Tables \ref{tab:fitsParametersNKfixedUnitary}  and  IV.  When $NK$ is small the  cubic term $b_3$ in the fitting polynomial is not needed to obtain a good fit, so in this case we leave it at zero.    

 Because these fits are successful,  the localization length's  $N \rightarrow \infty$ limit can be read off from the $b_0$ values in the left columns of Tables \ref{tab:fitsParametersNKfixedUnitary}  and  IV.  For $NK = 1,2,3$, and for both unitary and orthogonal ensembles, $\xi$'s $N \rightarrow \infty$ limit lies between $0.25$ and $0.45$.  % This is numerical proof of the presence of a localized phase in the site-diagonal-disorder model in one dimension when $N\,K$ is kept fixed while taking the $N \rightarrow \infty$ limit.

%   Unlike the $K$ fixed limit where $\xi$ decreases when $N$, in the $NK$ fixed limit $\xi$ decreases monotonically with $N$.   
    % Our results also show that corrections to the $NK$  fixed limit are controlled by powers of $N^{-1/2}$.  

\begin{table}
% These were taken from fittingParameters.xls, under localizationLength.
  \centering 
   \label{tab:fitsParametersNKfixedUnitary}
  \caption{Fitting Coefficients, $NK$ fixed, Unitary Ensemble. The data that was fitted had statistical errors of about 0.2\%.}
  \begin{tabular}{|c|c|c|c|c|clclclcl}
\hline
$NK$ & $b_0$ & $b_1$ & $b_2$ & $b_3$ \\ \hline
1&0.292736&2.2361  &2.8857&-  \\ \hline
2&0.287002&4.97338&3.28302	&38.3808 \\ \hline
3&0.413194&4.49159&29.6528	&76.9259 \\ \hline
\end{tabular}
\end{table}

\begin{table}
% These were taken from fittingParameters.xls, under localizationLength.
  \centering 
   \label{tab:NKfixedOrthogonal}
  \caption{Fitting Coefficients, $NK$ fixed, Orthogonal Ensemble}
  \begin{tabular}{|c|c|c|c|c|clclclcl}
\hline
$NK$  & $b_0$ & $b_1$ & $b_2$ & $b_3$ \\ \hline
1& 0.279892&2.09338&1.42885	& - \\ \hline
2&0.360778&2.81914&11.4255	& - \\ \hline
3&0.400219&3.91652&24.2491&11.377\\ \hline
\end{tabular}
\end{table}

  \section{Discussion of the Numerical Results}
These results are a numerical proof that  the one-dimensional site-diagonal-disorder model is localized in the double scaling limit where $NK$ is kept fixed and $N$ is taken to infinity.  This localized property is expected of one-dimensional wires with short-range uncorrelated disorder and short-range hopping.  Wires in this class have been shown both numerically and analytically to be localized, except in the special case of certain ensembles with perfectly conducting channels.  \cite{ishii1973localization,kramer1993localization}

These numerical results demonstrating a localized phase in the site-diagonal-disorder model are in contrast with Wegner's proof that the same model is always conducting.  Although there is a contrast, there is no contradiction: Wegner's proof concerns the  $N \rightarrow \infty$, $K$ fixed limit, which removes the physics that is responsible for localization. 

 These results also  suggest that in three dimensions the site-diagonal-disorder model may exhibit both localized and conducting phases, and an Anderson transition between them. If so then its behavior would be substantially the same  as the three-dimensional Anderson model.

 \section{The Gauge Invariant Model}

  In order to go beyond one dimension, and to obtain non-numerical insight, we will now analyze both the $K$ fixed limit and the $N\,K$ fixed limit using analytical analysis of a functional integral. In order to ease  our mathematical analysis, we will switch from Wegner's site-diagonal-disorder model to Wegner's gauge invariant model, both of which were introduced in the same paper.  \cite{PhysRevB.19.783}  The only difference between the two models is that in the gauge invariant model the hopping between sites is mediated by random matrices, while in the site-diagonal-disorder model hopping is mediated by deterministic matrices proportional to the identity.   We will use however a more flexible notation for the hopping, which is general to any geometry, connectivity, or dimension.
  
  Like the site-diagonal-disorder model and the Anderson model, the gauge invariant model is a tight-binding Hamiltonian; i.e. the electrons live on a lattice or graph geometry with $V$ sites and $N$ electron orbitals at each site.  There are $N$ basis elements at each site, so the total basis size is $N \times V$. We will use the lower case letters $n, v$ respectively to denote the orbital index and the position. 

We develop the unitary ensemble version of the gauge invariant model, where the Hamiltonian is hermitian.  This work can easily be generalized to the orthogonal ensemble\cite{Fyodorov02, Fyodorov02a} and probably also to the symplectic ensemble. In the orthogonal ensemble variant the spin variables occuring in the functional integral are  $4 \times 4$ matrices, as opposed to the $2 \times 2$ matrices which we will encounter here.

The hamiltonian $\tilde{H}$, including both its on-site and hopping parts, is random and fully described by the second moment 
  \begin{eqnarray} \label{DisorderSecondMoment}
  &\;& {\overline{{\tilde{H}_{n_{1} n_{2} v_{1} v_2}} {\tilde{H}_{n_{3} n_{4} v_{3} v_4}} }} \nonumber \\
  &=&  { \tilde{\epsilon}}^2 {{N}^{-1} {(1-\kappa)_{v_1 v_2}} {\delta_{n_{1} n_{4}}} {\delta_{n_{2} n_{3}}} {\delta_{v_{1} v_{4} }} {\delta_{v_{2} v_{3} }} }
  \end{eqnarray}
     The system geometry, including the number of dimensions and all other structural details, is encoded in the positive indefinite operator $\kappa$, which is the only nonlocal operator in the model.  $\kappa$ controls kinetics and hopping, and its norm is $K$. We require that the $1-\kappa$   operator be positive definite in order to assure convergence of a Hubbard-Stratonovich transformation used to derive the functional integral representation of this model.   We  require that $\kappa$  be  a Laplacian, meaning that $\kappa |\vec{0}\rangle = 0$, where $|\vec{0}\rangle$ is the spatially uniform vector.  The requirement that $\kappa$ be a Laplacian is physically the minimal requirement for it to be able to capture the physics of hopping and some kind of graph or lattice geometry, and is mathematically required to capture the global continuous symmetry which is key to the physics of conduction and localization.  It can generally be understood that $\kappa$ is small compared to $1$, though we do not require this except when taking the $N \, K$ fixed, $N \rightarrow \infty$ limit. 
     %todo: do we need this?  We also require that $1 > \langle \vec{s} | k|\vec{s}\rangle > 0 \; \forall \; \vec{s} \neq 0$, where $ |\vec{s}\rangle$ are eigenvectors with momentum $\vec{s}$.  
     $\tilde{\epsilon}$ is the energy scale of  the Hamiltonian.

The site-diagonal-disorder model and the gauge invariant model are very much alike.   The main physical difference between the two models is in the scattering length, which is zero (on-site) in the gauge invariant model, but can have any value in the site-diagonal-disorder model.   In other words, the gauge invariant model  is an effective field theory: it  describes the physics of diffusive conduction and of localization at scales longer than the scattering length, but omits ballistic physics at scales shorter than the scattering length.   There are two notable mathematical simplifications in the gauge invariant model. The first is that its energy band is the semicircular one of random matrix theory, while the site-diagonal-disorder energy band can have any shape at all, according to the kinetic operator (hopping matrix) which one chooses. The second is that the gauge-invariant model's long distance physics, i.e. movement of its cooperon and diffuson,  is directly controlled by the kinetic operator $\kappa$, which is an input parameter.  In the site-diagonal-disorder model the kinetics of the cooperon and diffuson is regulated by more complicated mathematics.  

 \section{Functional Integral Formulation of the Gauge Invariant Model}
 
 The conversion of disordered models to functional integrals is a well developed topic that originated with Schafer and Wegner in 1980. \cite{Schafer80,PhysRevLett.46.490}    The main point of these functional integrals is that their spin variables obey a continuous symmetry and therefore are capable of a transition from a spontaneously-broken-symmetry phase to a symmetric phase.  In the symmetry-broken phase the spins are correlated over long distances, which corresponds to electronic conduction over long distances.  In the symmetric phase spins  have a finite correlation length, so when two spins are separated by a distance longer than that length their correlator is exponentially small.  This symmetric phase corresponds to the Anderson localization phase, where conduction  over distances longer than the localization length is exponentially small. The correlation length of the spins is the same as the localization length.  Thus functional integrals bring to light the spontaneous symmetry breaking that lies at the core of the phase transition from Anderson localization to long-distance conduction.
 
A general feature of these functional integrals  is that the conversion from disordered Hamiltonians to functional integrals can be accomplished if desired with mathematical rigor, i.e. there is a strict equivalence between the original ensemble of Hamiltonians and the seemingly much different functional integral.  Another general feature is that the spin variables in the functional integrals are always matrix variables, and that the matrices must be decomposed into eigenvalues vs. rotations.  The rotations possess the global continuous symmetry at the heart of the Anderson phase transition, while the eigenvalues always remain massive, reflecting the short-distance physics associated with the scattering length.

Although these features are universal, for each disordered model there are several alternative functional integrals.  The oldest one is based on the replica technique, which is more flexible but relies on the replica trick where the number of replicas is taken to zero.  \cite{PhysRevLett.46.490,kamenev1999wigner} The other alternative functional integrals generally involve graded matrices, i.e. matrices  in which half the variables are anticommuting (Grassmann) variables and the other half are the more commonly known commuting variables.  The first such integral was developed by Efetov, who used the Hubbard-Stratonovich transformation to derive the functional integral, adapting that transformation to graded matrices. \cite{Efetov82a,Efetov97} More recently superbosonization was introduced, which avoids the Hubbard-Stratonovich transformation in favor of a mathematical step of pushing one integration domain onto a more restricted domain of graded matrices.  \cite{bunder2007superbosonization,PhysRevB.96.054208} 

Here we will use  a hybrid approach pioneered by Fyodorov.  \cite{Fyodorov02, Fyodorov02a} This technique judiciously combines both a Hubbard-Stratonovich transformation and a push. Its principal advantage is that half of the variables - the anticommuting variables - can be integrated exactly.  The result of this integration is that the resulting functional integral contains two different matrix spins, plus a determinant which is a direct result of the integration.  While this form is a bit more complicated than alternative functional integrals, it does have the advantage that the remaining variables in the model are all commuting, which simplifies their mathematical treatment.  It should be noted that a very similar procedure resulting in a functional integral with a determinant in the place of Grassmann variables had already been performed twenty years earlier by Ziegler and developed by other authors. \cite{Ziegler82,Brezin85,Constantinescu87, Disertori02, disertori2010quasi} However Ziegler confined his work to a functional integral suitable for calculating the density of states, while Fyodorov distinguished himself by producing a Grassman-free functional integral suitable for calculating the advanced-retarded two point correlator $R_2^{AR}$.

Probably any of the competing functional integrals can be analyzed in terms of energy and length scales to give insight into the absence of a localized phase in the $N$ fixed, $N \rightarrow \infty$ limit and the presence of that phase in the $N \, K$ fixed limit.  We adopt the hybrid approach here for reasons of personal experience and taste, and to showcase certain qualities which the hybrid approach offers: clear mathematics without need for modern results on graded matrices, and a very systematic if rather complicated arithmetic and perturbation theory.  Moreover, because  variables can be grouped together and manipulated as commuting matrices, calculation of the leading-order value of the advanced-retarded density-density correlator $R_2^{AR}$ proceeds with very quickly, with a minimum of algebra, and without any restrictions on the level spacing parameter $\omega=E_1-E_2$; we supply a sketch of this calculation in the appendix. This correlator is a well-known result that has been calculated using a variety of approximations and techniques. \cite{Altshuler86,Andreev95, Andreev96,Kravtsov94,Mae03,Takahashi04}

Fyodorov developed the hybrid approach for zero-dimensional (single-site) models. \cite{Fyodorov02, Fyodorov02a}  To perform the push step in deriving the functional integral, Fyodorov used an integral that was first done  by David, Duplantier, and Guitter, \cite{David93} and that was later connected to  the Singular Value Decomposition. \cite{Spencer04}   Disertori generalized Fyodorov's hybrid approach to the gauge invariant model in extended systems. \cite{DisertorisModel}   Using Fyodorov's hybrid approach, Wegner's gauge invariant model   transforms in a completely mathematically rigorous way  into the following functional integral: \footnote{  References \onlinecite{Fyodorov02, Fyodorov02a} explain how to perform the hybrid transformation in the zero dimensional case.  For a detailed and rigorous explanation of how to perform the  hybrid transformation in any dimension for both the gauge invariant model and the site-diagonal-disorder model, please see Equations 1 to 21 in my preprint, Ref. \onlinecite{sacksteder2009mathematically}.  Although that preprint contains numerous mistakes, the derivation of the hybrid functional integral in Equations 1 to 21 is rigorous and correct.  The "Disertori model" that I refer to in that preprint is Wegner's gauge invariant model, and the Wegner model that I refer to is Wegner's site-diagonal-disorder model. }
% If in addition the source $J$ is  diagonal in the position index $v$ (restricting the theory to calculating  on-site elements of the Green's function $G_{vv}$) then  $A^0$ is local and $\det A^0$ factorizes site by site.  $A^1$ simplifies to $  {Q^b_{v_1}} {L_{j_{1}}} {(1-k)_{v_1 v_2} {{\delta}_{i_1 i_2}}}$, so that there is no dependence on $W$; the remaining integral in $e^{\mathcal{L}_{eff}}$ is just $(\int dW)^V$.  
%Adjusting  for factors of two caused by Fyodorov's use of the integration measure ${dS}{dS^*}$ versus our use of ${dS^R}{dS^I}$, the integral  is $\int {dW} = { {{(\prod_{l=N-2}^{N-1}{l!})}^{-1}}  {\pi}^{{-1} + 2N} }$.  When $J^f =0 $ we obtain the following form:
\begin{eqnarray}
{\bar{Z}} & = & {\gamma {\int_{Q^b \geq 0} dQ^f \; dQ^b\; \;   e^{\mathcal{L}}  }}  \nonumber \\
& \times & {\det({ Q^f_{v_1}  {\delta_{v_1 v_2}} {\delta_{j_1 j_2}} } - {  Q^b_{v_1} L  (1 - \kappa)_{v_1 v_2} {\delta_{i_1 i_2}}  }   )} 
\nonumber \\
{\mathcal{L}}  & = &  {(N - 2) \sum_v {Tr}(\ln Q^f_v)} + {(N - 2) \sum_v {Tr}(\ln Q^b_v L)} 
\nonumber \\
& + & {{\imath N}{\tilde{\epsilon}}^{-1} \sum_v {Tr}(Q^b_v L  ({\hat{E}} - J^b_v)  )} 
+ {{\imath N}{\tilde{\epsilon}}^{-1} \sum_v {Tr}(  Q^f_v  {{\hat{E}}} )}
 \nonumber \\ 
 & - & { \frac{N}{2} \sum_{v_1 v_2} {(1-\kappa)^{-1}_{v_1 v_2}}{Tr}{(Q^f_{v_1}  Q^f_{v_2}  )} } 
 \nonumber \\
 & - & {\frac{N}{2}{{\sum_{v_1 v_2} {{(1-\kappa)_{v_1 v_2}} {Tr}{(Q^b_{v_1} L Q^b_{v_2} L  )}}}}}
\nonumber \\
\gamma & = & {  N^{{{2 N V } } + {2  V }}  { {((N-2)! (N-1)! )}^{-V}}  }
\nonumber \\
& \times & {2^{ - { V }}   {\pi}^{{-3V} }   } {(\det (1-\kappa))}^{-{2}} 
 e^{{\frac{NV}{2 {\tilde{\epsilon}}^2} {Tr}({\hat{E}} {\hat{E})}}}   
\label{DisertoriModel1}
\end{eqnarray}
The degrees of freedom $Q^f$ and $Q^b$ are $2 \times 2$ Hermitian matrices, and the logarithms and multiplications are matrix operations.  The ${Q^b \geq 0}$  means that $Q^b$ is constrained to have only positive or zero eigenvalues;  the integration measure contains a theta function $\theta(Q^b)$.  The $L$ matrix is equal to the Pauli $\sigma_z$ matrix; it is diagonal with $+1$ and $-1$ eigenvalues.    The $\hat{E}$ operator contains the energies at which the densities are measured; it is diagonal with eigenvalues $E_1$ and $E_2$.  The source matrices $J^b_v$ are also diagonal matrices, with two entries on the diagonal.   The matrix ${ Q^f_{v_1}  {\delta_{v_1 v_2}} {\delta_{j_1 j_2}} } - {  Q^b_{v_1} L  (1 - \kappa)_{v_1 v_2} {\delta_{i_1 i_2}}  }  $ in the determinant lives in a $2 \times 2 \times V$ space - the $V \times V$ matrix $\kappa$ mediates transitions between positions $v_1, v_2$, the $2 \times 2$ matrix  $Q^f$ mediates transitions between $i_1,i_2$, and the $2 \times 2$ matrix  $Q^b$ mediates transitions between $j_1,j_2$. \footnote{If one starts with  equation \ref{DisertoriModel1} and fixes the eigenvalues of $Q^f, \, Q^b$ to their saddle point values (which results in the disappearance of the logarithm terms in the Lagrangian) , then the resulting Lagrangian is sufficient to reproduce all  results in Muzykantskii and Khmelnitskii’s article on Anomalously Localized states. \cite{Muzykantskii95} This is shown in Ref. \onlinecite{sacksteder2009mathematically}'s subsection on Anomalously Localized states.  }

The ensemble-averaged value of the density $\rho(v_1,E_1)$ at site $v_1$ and energy $E_1$ can be calculated by taking the first derivative of ${\bar{Z}}$ with respect to $J^b_{v_1,j=1}$ and then setting $J^b=0$. Equivalently, one can just insert a factor of $-\imath N  {\tilde{\epsilon}}^{-1} (Q^b)_{v_1, j=1}$ into the body of the functional integral and then set $J^b=0$. 

Taking this a step further, the Advanced-Retarded correlator $R_2^{AR}$, which gives the correlation of the density $\rho(v_1,E_1)$ at site $v_1$ with the density $\rho(v_2,E_2)$ at site $v_2$, can be calculated by  using the functional integral to obtain the average value of $- N^2  {\tilde{\epsilon}}^{-2} (Q^b)_{v_1, j=1} (Q^b)_{v_2, j=2}$. The Advanced-Retarded correlator $R_2^{AR}$  reveals whether  $Q^b$'s values  are correlated across long distances, which is the case in the conducting phase.  Alternatively  $R_2^{AR}$ can show that  $Q^b$'s correlations die off exponentially with distance, which is the case in the localized phase.  Another way of saying this is that if $Q^b$ is everywhere pinned to the same value then we have conduction, while if instead $Q^b$ fluctuates from point to point then we have Anderson localization.

The development and final form of this functional integral are strongly constrained by the fact that the disorder in the original gauge invariant model is static; it does not depend on time.  As a consequence, the spin variables in the functional integral have important relationships with each other; the functional integral possesses important symmetries.  In alternative functional integrals for the same gauge invariant model there is only one matrix, either living in a space  of replicas, or in a graded  space containing both commuting and anticommuting (Grassman) variables.  In the  hybrid model adopted here the $Q^b$ and $Q^f$ matrices are analogous to the commuting variables in the supersymmetric formulation, and the determinant figuring so prominently in the functional integral is the result of performing an exact integration of the anticommuting variables.

One spectacular consequence of the functional integral's symmetry is that if one does not take any derivatives with respect to the sources, i.e. if one does not average over $Q^b$, then the integral's value is identical to one.  This result is not immediately apparent in Equation \ref{DisertoriModel1} and  a perturbative analysis does not give a clear  indication of this result.  In order to give the correct result (one), at each order of perturbation theory large numbers of diagrams must sum to zero.    Part of the reason why perturbation theory does not easily reveal this result is that the $Q^f$ and $Q^b$ matrices are given different roles; for instance the ${Tr}(Q^f_{v_1}  Q^f_{v_2}  ) $ term is multiplied by $(1-\kappa)^{-1}$ while the ${Tr}{(Q^b_{v_1} L Q^b_{v_2} L  )}$ term is multiplied by $(1-\kappa)$. However the symmetry between $Q^f$ and $Q^b$ can be partially restored by performing nonlocal transformations of $\overline{Q}^f = (1-\kappa)^{-1/2} Q^f$ and $\overline{Q}^b = (1-\kappa)^{1/2}Q^b$, arriving at an alternative formulation of the same functional integral:

\begin{eqnarray}
\bar{Z} & \propto & { {\int d\overline{Q}^f \; d\overline{Q}^b\; \;   e^{\mathcal{L}}  }}  
\nonumber \\
& \times & {\det({ \overline{Q}^f_{v_1}  {(1 - \kappa)^{-1/2}_{v_1 v_2}} {\delta_{j_1 j_2}} } - {  \overline{Q}^b_{v_1} L  (1 - \kappa)^{1/2}_{v_1 v_2} {\delta_{i_1 i_2}}  }   )} 
\nonumber \\
& \times & \Pi_{v_1} \theta( \sum_{v_2} (1-\kappa)^{-1/2}_{v_1 v_2} \overline{Q}^b_{v_2} L )
\nonumber \\
\mathcal{L}  & = &  {(N - 2) \sum_{v_1} {Tr}(\ln ( \sum_{v_2} (1-\kappa)^{1/2}_{v_1 v_2} \overline{Q}^f_{v_2}))} 
\nonumber \\
 & + &   {(N - 2) \sum_{v_1} {Tr}(\ln( \sum_{v_2} (1-\kappa)^{-1/2}_{v_1 v_2} \overline{Q}^b_{v_2} L))} 
\nonumber \\
& + & {{\imath N}{\tilde{\epsilon}}^{-1} \sum_v {Tr}(\overline{Q}^b_v L  {\hat{E}}   )} 
+ {{\imath N}{\tilde{\epsilon}}^{-1} \sum_v {Tr}(  \overline{Q}^f_v  {{\hat{E}}} )}
 \nonumber \\ 
 & - & \sum_v  { \frac{N}{2}  {Tr}{(\overline{Q}^f_{v}  \overline{Q}^f_{v}  )} } 
- \sum_v {\frac{N}{2}{{ { {Tr}{(\overline{Q}^b_{v} L \overline{Q}^b_{v} L  )}}}}}
\nonumber \\
& - & {{\imath N}{\tilde{\epsilon}}^{-1} \sum_{v_1 v_2} (1-\kappa)^{-1/2}_{v_1 v_2} {Tr}(\overline{Q}^b_{v_1} L  J^b_{v_2}  )} 
\label{SymmetrizedModel}
\end{eqnarray}
After this transformation the principal formal difference between $\overline{Q}^f$ and $\overline{Q}^b$ is that $\overline{Q}^b$'s rotations live on a hyperbolic manifold that is not compact, while $\overline{Q}^f$'s rotations live on a half-sphere manifold that is compact.  The two manifolds can be parameterized in a way that is identical except for a difference in signs, i.e. a plus sign in one manifold's parameterization vs. a minus sign in the other manifold's parameterization.  Similarly there is a difference in signs in the kinetic terms now embedded inside of the logarithms and determinant.  At the level of the saddle point approximation there is an additional difference: $\overline{Q}^b$'s two eigenvalues at the saddle point are of the same sign, while  $\overline{Q}^f$'s two eigenvalues are of opposite signs.  In consequence,  at the level of perturbation theory around the saddle point the only differences between $\overline{Q}^b$ and $\overline{Q}^f$ are certain minus signs.

\section{Two Energy and Length Scales}

In this section we will use the hybrid functional integral in Eq. \ref{SymmetrizedModel} to show  that the gauge invariant model has two natural energy scales.  The first is the scattering energy scale $\tilde{\epsilon} N$. The second is the much smaller energy scale $\tilde{\epsilon} N K$ which controls conduction and localization.  Corresponding to these two energy scales are two length scales: the scattering length which is equal to 1 in the gauge invariant model because scattering occurs on site, and the localization length which can be much larger or even diverge.   The beauty of the functional integral approach to disordered models is that these two scales can be separated from each other by breaking the matrix variables into two sectors: their eigenvalues and their rotations. 

We will first analyze the  eigenvalues of $\overline{Q}^f, \overline{Q}^b$, which   mediate scattering physics.  Their saddle-point values are of order $O(1)$, except near the band edges where the saddle-point approximation breaks down.  In contrast, the eigenvalues' fluctuations are small, of order $O(1/\sqrt{N})$, and do not show significant inter-site correlation. The net effect is that the eigenvalues are basically the same at every site in the system. 

We will next analyze the rotations of $\overline{Q}^f, \overline{Q}^b$, which mediate conduction and localization physics.  In a conducting system these rotations maintain the same value (except for small fluctuations) throughout the entire system, and $\overline{Q}^f, \overline{Q}^b$ retain the same value everywhere. In an insulating system the fluctuations in the rotations grow steadily  with distance  and reach $O(1)$ at the localization length scale.  In the gauge invariant model the kinetic term which penalizes fluctuations in the rotations is proportional to $N\,K$.  Therefore  in the $N \rightarrow \infty$ limit with $K$ (the magnitude of the kinetic operator $\kappa$)  fixed the kinetic term diverges.  The divergence forces the fluctuations  to zero, forcing the system into the conducting phase.  In contrast, when $N\,K$ is kept fixed the kinetic term does not diverge. In this case, because the kinetic operator $\kappa$ is a Laplacian with a zero eigenvalue when acting on spatially constant rotations, it has small eigenvalues when acting on long-wavelength rotations.  It therefore can fail to  eliminate rotations over long length scales, resulting in the localized phase.  This is the key reason why the $N$ fixed limit eliminates the localized phase while the $N \, K$ fixed limit retains it.

Lastly we will analyze the determinant.  This contains two sectors: one sector with eigenvalues proportional to $N$, and another with eigenvalues proportional to $N \, K$.  Because $\kappa$ has a zero eigenvalue, the latter sector includes two very small eigenvalues signalling the same physics as the globally uniform rotations of  $\overline{Q}^f, \overline{Q}^b$.  The two sectors are exact analogues of the  eigenvalue and rotation sectors of the $\overline{Q}^f, \overline{Q}^b$ matrices.

\subsection{About the style of the present analysis of the functional integral}
 The style in this present work of our analysis  of the functional integral  is not rigorous.  A fully systematic and maximally rigorous analysis  would start with  careful separation of globally uniform rotations from other rotations where some sites rotate differently from others, and would then perform a nonperturbative  integration of the globally uniform rotations.  Maximal rigor would also require a detailed solution of the saddle point equation including all terms that contribute to it, a systematic perturbative expansion of the determinant, and a systematic perturbative expansion of the  corrections to the saddle point, including fluctuations of both the  eigenvalues and the rotations. \footnote{The mathematical structure of perturbative corrections to the saddle point is quite sensitive to the exact choice of which terms are included and which terms are left out in the saddle point equations. In order to obtain a perturbation theory where no diagram diverges in the the $N \rightarrow \infty,\; N\,K$ fixed limit, it is necessary to include in the saddle point equations terms where the eigenvalues couple to fluctuations in the angular variables.  As a result of this inclusion the saddle point values of the eigenvalues fluctuate from point to point on the lattice,  with a magnitude that is controlled by powers of $N^{-1/2}$.   } These requirements are all within reach, except for two difficulties.  The first difficulty is that it is mathematically questionable to make a perturbative expansion of fluctuations in the rotations because those fluctuations are not around a single value of the rotation matrix but instead around all points on a manifold of spatially uniform rotations.  This  difficulty manifests itself in divergences of the Jacobian which is produced when separating the  globally uniform rotations from the other rotations.  Further details about the Jacobian and its divergence are included in Appendix \ref{JacobianAppendix}.
 
   A second difficulty is that the gauge invariant model has competing saddle points,  and the choice of saddle point can in principle be different at each site on the lattice.  Unless the globally uniform saddle points are strongly favored over their alternatives where some sites pick one saddle point and other sites pick another, perturbative analysis of the functional integral is impossible because of the exponential proliferation of saddle points, each of which violates translational invariance.     This issue was considered in two rigorous proofs about certain functional integrals used to study the density of states in disordered models \cite{Constantinescu87,Disertori02}.  Such functional integrals used to study the density of states are structurally simpler than the integrals used to study  the Advanced-Retarded two point correlator $R_2^{AR}$, but unfortunately   the density of states does not contain information about the localized and conducting phases, while $R_2^{AR}$ does.   In these two proofs it was shown that a single global saddle point dominates, and also  rigorous control over fluctuations around the saddle point was obtained using coupled cluster expansions.  To the author's knowledge, no similar effort has been taken to control non-uniform saddle points in functional integrals built for computing the Advanced-Retarded two point correlator $R_2^{AR}$.    Both of these two difficulties  are of a non-perturbative nature, so a narrow focus on purely perturbative mathematics will naturally bypass them.

 Leaving these two difficulties aside, a maximally rigorous analysis of the saddle point equations and the perturbative corrections to the saddle points results in a mathematical structure with  more than two dozen different vertices that combine to produce a wide variety of  Feynman diagrams.   While there is nothing particularly difficult in principal about this mathematics, the huge variety of vertices and diagrams is rather difficult to manage.  It is possible to classify all vertices and Feynman diagrams  in powers of $N$, or alternatively in powers of $N\,K$.  It is easily possible to show  that in the $K$ fixed limit there is a finite and very small number of diagrams which survive the $N\rightarrow \infty$ limit; in other words this limit is trivial which is why it is conducting.  With a great deal more effort it is possible to determine a subset of vertices which survive  the $N\rightarrow \infty, \; N\,K$ fixed limit, and (if the saddle point equations were formulated and solved with sufficient care) to show that there is no diagram which diverges in this limit.  \footnote{ Ref. \onlinecite{sacksteder2009mathematically} made an attempt at a maximally rigorous research program, but included many minor and substantial errors.   Further work to remedy those errors resulted only in the extremely restricted results alluded to in these paragraphs.  The extreme complexity of the perturbation theory discourages work on, for instance, calculating next-to-leading order corrections. In particular, the author's original goal of supplying a new level of mathematical rigor in understanding the conducting phase (in the $N\,K$ fixed limit where that phase is not trivial) appears to be unreachable. } The fact that there are several vertices that remain finite in this limit and that they combine to produce an infinite number of Feynman diagrams is a perturbative indication that the $N\,K$ fixed limit produces a sigma model which in principal is capable of producing a localized phase.  
 
In the present work we leave aside attempts at such rigor and move less systematically.  We  look broadly at the scales in play in order to show that some sectors of the functional integral are controlled by $N$ while others have kinetics that are controlled by $N \, K$.  This is sufficient to provide an account of why the $N$ fixed limit removes the localized phase while the $N \, K$ fixed limit preserves it. 
%  Scattering occurs on-site.  In the conducting phase fluctuations in the density of states have a characteristic correlation length which is short in good conductors and diverges at the Anderson phase transition.  In the localized phase the corresponding length scale is the localization length, which is short in strongly disordered materials and diverges at the Anderson phase transition.    Corresponding to the scattering length and the correlation/localization length, there are two energy scales. The energy scale of scattering is $N$, while the energy scale regulating  correlations in the density of states is $N K$, a much smaller value.    
%In the conducting phase the two energy and length scales can be disentangled by separating the $Q^f$ and $Q^b$ matrices into eigenvalues and rotations.  The eigenvalues of  $Q^f$ and $Q^b$ control the scattering physics, while rotations control conduction and localization.  \

\subsection{The Eigenvalues}
We begin our analysis of the functional integral by decomposing the matrix spins  $\overline{Q}^f,\,\overline{Q}^b$ into eigenvalues and rotations.  For $Q^f$ the correct change of variables is $\overline{Q}^f_v = U_v x^f_v U_v^\dagger$, where the eigenvalue matrix $x^f_v$ is  diagonal, the rotation $U_v$ is unitary, and $v$ is the position index.   The new integration measure is $d\overline{Q}^f =   \Delta_{VdM}^2(x^f) \, {dU} \, {dx^f} $ where $\Delta_{VdM}(x^f) = x_{1}^f - x_{2}^f$ is the Van der Monde determinant.  In the previous sentence the subscript of $x^f$ distinguishes the two eigenvalues at position $v$.    Note that the unitary matrix $U$ lives on the $U(2)$ manifold which is compact - it can be parameterized with variables that are all bounded.

The correct change of variables for $\overline{Q}^b$ requires a little more care because of the constraint that $\overline{Q}^b$ must be positive indefinite; $\overline{Q}^b \geq 0$.  It is also important that $\overline{Q}^b$ is always paired with $L=\sigma_z$.
%, which is a consequence of our use of  $L$ to make the $dS$ integrals converge. 
  Fyodorov \cite{Fyodorov02} showed  that $\overline{Q}^bL$ factors into $\overline{Q}^b_v L = T_v x^b_v T^{-1}_v$, where $x^b$ is diagonal and constrained by $x^b L \geq 0$,  and $T$ is a member of the pseudo-unitary hyperbolic group $U(1, 1)$ defined by $T^\dagger L T = L$.   The integration measure is $d\overline{Q}^b = dx^b \, dT \, \Delta^2_{VdM}(x^b)$.   Note that $T$ lives on the  $U(1,1)$ manifold which is not compact and is indeed hyperbolic. As a consequence  $T$ has parameterizations in which one or more of its parameters is unbounded.  The presence of unbounded degrees of freedom in the model is an important mathematical feature which can complicate analysis.

The behavior of the eigenvalues can be most easily understood by leaving out the rotations, so that $\overline{Q}^f \rightarrow x^f$ and $\overline{Q}^b \rightarrow x^b$. The Langrangian for $x^f$ (that for $x^b$ is similar) is  
\begin{eqnarray}
 \mathcal{L}  & = &{(N - 2) \sum_{v_1} {Tr}(\ln ( \sum_{v_2} (1-\kappa)^{1/2}_{v_1 v_2} x^f_{v_2}))} 
\nonumber \\
& + &  {{\imath N}{\tilde{\epsilon}}^{-1} \sum_v {Tr}(  x^f_v  {{\hat{E}}} )} - \sum_v  { \frac{N}{2}  {Tr}{(x^f_{v}  x^f_{v}  )} } 
 \end{eqnarray}
 All of these terms are proportional to $N$, so that as $N$ increases toward the $N \rightarrow \infty$ limit $x^f$ becomes progressively more closely pinned to its saddle point value.  Expanding to second order in the fluctuations of $x^f$ around the saddle point immediately reveals that they are of order $O(N^{-1/2})$  and that their second moment is of order $O(N^{-1})$.  Compared to these quite small on-site fluctuations, the correlations between fluctuations at different sites are even smaller: they are suppressed by factors of $\kappa$, which is small compared to $1$. 
 
  Concerning the saddle points themselves, there are two such saddle points that give leading contributions, which we will label as the $+$ saddle point and the $-$ saddle point.   At both  saddle points the $x^b$ values are the same:
 \begin{eqnarray} 
 x^b_1 =  \sqrt{1 -(E_1/2\tilde{\epsilon})^2} + \imath E_1/ 2 \tilde{\epsilon}
 \nonumber \\
 x^b_2 =  \sqrt{1 -(E_2/2\tilde{\epsilon})^2} + \imath  E_2/ 2 \tilde{\epsilon}
 \end{eqnarray}
At the $+$ saddle point the $x^f$ values are the same as the $x^b$ ones i.e. $x^f_1 = x^b_1$ and $x^f_2 = x^b_2$, while at the $-$ saddle point they are interchanged i.e. $x^f_1 = x^b_2$ and $x^f_2 = x^b_1$. In addition there are two suppressed saddle points where the sign of $x^f_1$ or $x^f_2$ is reversed.

\subsection{Global Rotations}
It  is necessary to make a distinction between on one hand  global rotations where every site in the entire system rotates in synchrony, and on the other hand non-uniform rotations where the sites do not all rotate together.   This is necessary because, when $E_1$ is set equal to $E_2$, i.e. when $\hat{E}$ is proportional to the identity, the integrand inside the  functional integral is invariant under global rotations.  Therefore the global rotations cannot be performed perturbatively, and must be done exactly.  In contrast the non-uniform rotations may be integrated using pertubation theory.  Therefore a change of variables must be performed, from the original rotation variables which are each located at individual sites, to a set of rotation variables where the global rotation is distinct and separate from all other rotation variables.  Accompanying this change of variables, a Jacobian must be introduced to the functional integral.  As mentioned before and discussed in Appendix \ref{JacobianAppendix}, the Jacobian diverges
 %produces powers of the inverse energy splitting $\omega^{-1} = (E_1-E_2)^{-1}$ 
 at certain non-zero points on the manifold of global rotations.  Excepting this divergence, its contributions to perturbation theory are weighted by a factor of the inverse system volume $V^{-1}$, making them rather small.  However, because of the divergence, the Jacobian developed in Appendix \ref{JacobianAppendix} can be used only if the energy splitting $\omega$ is larger than the level spacing $\Delta E$, in which case the manifold of global rotations is not fully explored.  We neglect the Jacobian in the remaining discussion. 

 Fortunately, in the absence of the Jacobian, the  integrals over global rotations can be done exactly. The integral over global rotations of $Q^f$ is the famous Harish-Chandra-Itzykson-Zuber integral\cite{HarishChandra56,Itzykson80}, and the  integral over global rotations of $Q^b$ is a well-known generalization of the same integral\cite{Fyodorov02a}.  The Harish-Chandra-Itzykson-Zuber integral is part of the math that produces the model's two saddle points, since they are related by a global rotation of $\overline{Q}^f$.   Once both integrals over global rotations have been performed, the average energy $(E_1+E_2)/2$ decouples from rotations and is coupled only to the eigenvalues $x^f,x^b$, influencing their saddle point values. There remains a coupling  of the energy splitting $\omega = E_1 - E_2$ to the rotations and to the source $J^b$, and the functional integral is multiplied by a factor $\omega^{-2}$.

Strictly speaking, it is necessary to perform the integration over global rotations prior to performing the saddle point approximation of the $x^b$ eigenvalues.  Rotations of $\overline{Q}^b$ lie on a non-compact manifold, so the Lagrangian's $\sum_v {Tr}(\overline{Q}^b_v L  {\hat{E}}   )$ term is unbounded.  For some points on the non-compact manifold of $\overline{Q}^b$'s rotations this pushes $x^b$'s saddle points outside of the band, making it impossible to perform the saddle point approximation.  The only correct way around this problem is to perform the exact integration over global rotations prior to performing the saddle point approximation.    Mae and Iida \cite{Mae03} and  Takahashi \cite{Takahashi04} were the first to address this  kind of issue and  perform the saddle point integration earlier than usual.   However this is a mathematical nicety that does not affect the remainder of our discussion here.

\subsection{Non-Uniform Rotations}
Next we examine all  other rotations besides the global rotation.  Since the eigenvalues remain very close to  their saddle point values and don't show any significant long distance behavior, we pin the eigenvalues to their saddle points and then analyze the rotations in isolation.   We set $\overline{Q}^f_v = U_v x^f_0 U_v^\dagger$ and $\overline{Q}^b_v L = T_v x^b_0 T^{-1}_v$, where $x^f_0,\, x^b_0$ are the saddle point values, and $U,\,T$ are the non-uniform rotations, i.e. the global rotations have been removed as described earlier.  Setting aside the source term, the functional integral simplifies to
\begin{eqnarray}
\bar{Z} & \propto & { {\int dU \; dT \;   e^{\mathcal{L}}  }}  
\nonumber \\
& \times & {\det({ \overline{Q}^f_{v_1}  {(1 - \kappa)^{-1/2}_{v_1 v_2}} {\delta_{j_1 j_2}} } - {  \overline{Q}^b_{v_1} L  (1 - \kappa)^{1/2}_{v_1 v_2} {\delta_{i_1 i_2}}  }   )} 
\nonumber \\
\mathcal{L}  & = &  {(N - 2) \sum_{v_1} {Tr}(\ln ( \sum_{v_2} (1-\kappa)^{1/2}_{v_1 v_2} \overline{Q}^f_{v_2}))} 
\nonumber \\
 & + &   {(N - 2) \sum_{v_1} {Tr}(\ln( \sum_{v_2} (1-\kappa)^{-1/2}_{v_1 v_2} \overline{Q}^b_{v_2} L))} 
\nonumber \\
& + & {{\imath N}{\tilde{\epsilon}}^{-1} \omega \sum_v {Tr}(\overline{Q}^b_v    )}  % possibly should have multiplied by omega/2 not omega?
+ {{\imath N}{\tilde{\epsilon}}^{-1} \omega \sum_v {Tr}(  \overline{Q}^f_v  L )}
%\nonumber \\
%& - & {{\imath N}{\tilde{\epsilon}}^{-1} \sum_{v_1 v_2} (1-\kappa)^{-1/2}_{v_1 v_2} {Tr}(\overline{Q}^b_{v_1} L  J^b_{v_2}  )} 
\label{SymmetrizedModelV2}
\end{eqnarray}
If there were no hopping, i.e. if the kinetic term $\kappa$ were equal to zero,  then this integrand would be invariant with respect to non-uniform rotations; the Lagrangian would simplify to a constant.  Therefore  the Lagrangian can be expanded in powers of $\kappa$:
\begin{eqnarray}
\mathcal{L}  & = & - {\frac{N - 2}{2} \sum_{v_1, v_2}\kappa_{v_1 v_2} {Tr}(\ln ( \overline{Q}^f_{v_1} \overline{Q}^f_{v_2} ))} 
\nonumber \\
 & + &   {\frac{N - 2}{2}  \sum_{v_1, v_2}\kappa_{v_1 v_2}  {Tr}(\ln(  \overline{Q}^b_{v_1} L \overline{Q}^b_{v_2} L))} 
\nonumber \\
& + & {{\imath N}{\tilde{\epsilon}}^{-1} \omega \sum_v {Tr}(\overline{Q}^b_v    )}  % possibly should have multiplied by omega/2 not omega?
+ {{\imath N}{\tilde{\epsilon}}^{-1} \omega \sum_v {Tr}(  \overline{Q}^f_v  L )}
%\nonumber \\
%& - & {{\imath N}{\tilde{\epsilon}}^{-1} \sum_{v_1 v_2} (1-\kappa)^{-1/2}_{v_1 v_2} {Tr}(\overline{Q}^b_{v_1} L  J^b_{v_2}  )} 
\label{SymmetrizedModelV2ExpansionInK}
\end{eqnarray}

%todo: introduce the parameterization:  Before performing the angular integrals we must choose a parameterization of $U(2)/U(1)\cdot U(1)$ and $U(1,1)/U(1)\cdot U(1)$.  

Many alternative parameterizations of $U$ and $T$ exist, but the following one seems to be particularly well suited for  performing a perturbation integration  of the non-uniform rotations:
\begin{eqnarray}
T &=& \begin{bmatrix} {\sqrt{1 + (y^b)^2 + (z^b)^2}} & {y^b - \imath z^b} \\ {y^b + \imath z^b} & {\sqrt{1 + (y^b)^2 + (z^b)^2}}  \end{bmatrix}, \nonumber \\
\;\; U &=&   \begin{bmatrix} {\sqrt{1 - (y^f)^2 - (z^f)^2}} & {\imath y^f + z^f}  \\ {\imath y^f - z^f}  & {\sqrt{1 - (y^f)^2 - (z^f)^2}} \end{bmatrix}
\label{PerturbativeParameterization}
\end{eqnarray}
%$U$ can be obtained from  $T$ with the substitution $y \rightarrow \imath y, \; z \rightarrow \imath z$.  The inverses of $U$ and $T$ can be obtained by inverting the sign of both $y$ and $z$.
$y^b$ and $z^b$ vary from $-\infty$ to $\infty$, while $y^f$ and $z^f$ vary within the unit circle defined by $(y^f)^2 + (z^f)^2 = 1$.  The  integration measures are $ {dy^f} {dz^f} $ and  $  {dy^b} {dz^b}$.   
These variables are convenient for perturbation theory because when the rotations $U,T$ are equal to the identity the $y,z$ variables are zero.  We will see that fluctuations in the $y,z$ variables are of order $(NK)^{-1}$.

%Because we made the change of variables  $Q^f_v = U_v x^f_v U^\dagger_v \rightarrow U_0 U_v x^f_v U^\dagger_v U^\dagger_0, \; Q^b_v L =  T_v x^b_v T_v^{-1} \rightarrow T_0 T_v x^b_v  T_v^{-1} T_0^{-1}$, we impose the constraints $0 = \sum_v \, y_v, \, z_v$.   

  After expanding in powers of $y, \; z$, the quadratic part of the  Lagrangian is:
  \begin{eqnarray} 
  \mathcal{L} & \approx &  {{\mp 2 \pi \imath N \omega \hat{\rho}} {\tilde{\epsilon}}^{-1}  \sum_{v }  ((y^f_{v})^2 + (z^f_{v})^2)}  
  \nonumber \\
&  + & {{2 \pi \imath N \omega \hat{\rho}} {\tilde{\epsilon}}^{-1}   \sum_{v }   ((y^b_{v})^2 + (z^b_{v})^2)  }  
%  + O(N \omega  \tilde{\epsilon}^{-1}   y^4 + N \omega  \tilde{\epsilon}^{-1}\tilde{x} y^2)
\nonumber \\
 & - & { \frac{8N}{2} \pi^2 \hat{\rho}^2 \sum_{v_1 v_2} \kappa_{v_1 v_2} (y_{v_1}^f y_{v_2}^f + z_{v_1}^f z_{v_2}^f)}
 \nonumber \\
 &-& {\frac{8N}{2} \pi^2 \hat{\rho}^2 {{\sum_{v_1 v_2} {{\kappa_{v_1 v_2} (y_{v_1}^b y_{v_2}^b + z_{v_1}^b z_{v_2}^b)} }}}} 
 %+ O( N  ky^4)
 \nonumber \\
2 \pi \hat{\rho} &= & \sqrt{1 - ( E /2 \tilde{\epsilon})^2}
\label{ExpressionsInYZCoordinates0}
\end{eqnarray}
Here N $\hat{\rho}$ is the gauge invariant model's density of states, and the $\pm$ sign specifies which saddle point is being used.  All four terms in the quadratic part of the Lagrangian are proportional to $N\, K$, which means that fluctuations in the $y,\,z$ angular variables  have a typical size which scales with $(N K)^{-1/2}$.  If $K$ is fixed during the $N \rightarrow \infty$ limit, then fluctuations in the angular variables are forced to zero; the angular variables have the same value at all sites on the lattice.  This is why the $K$ fixed $N \rightarrow \infty$  limit forces the gauge invariant model into the conducting regime.  

On the other hand,  taking the $N\,K$ fixed $N \rightarrow \infty$ limit leaves  the angular fluctuations with their magnitude unchanged.   Because $\kappa$ is a Laplacian, it acts very weakly on long  wavelength fluctuations.  On a regular  lattice the eigenvalues of $\kappa$ scale quadratically with momentum $\vec{k}$, i.e. as $|\vec{k}|^2$.  Therefore in the $N\,K$ fixed limit long-wavelength fluctuations can be very large.  This is why the $N\,K$ fixed limit is able to preserve the localized phase, while the $N$ fixed limit deletes it.

%The moments of $y, \, z$  are:
%\begin{eqnarray}
%\langle y_{v_1}^b y_{v_2}^b \rangle &=&   \langle z_{v_1}^b z_{v_2}^b \rangle = (8 N \pi^2 \hat{\rho}^2)^{-1} \langle v_1 |P_+  (\kappa - \imath ( 2 \pi \hat{\rho} \tilde{\epsilon})^{-1} \omega  )^{-1} P_+ | v_2 \rangle, 
%\nonumber \\
% \langle y_{v_1}^f y_{v_2}^f \rangle &=&  \langle z_{v_1}^f z_{v_2}^f \rangle =   (8 N \pi^2 \hat{\rho}^2)^{-1} \langle v_1 |P_+  (\kappa(1-\kappa)^{-1} \pm \imath ( 2 \pi \hat{\rho} \tilde{\epsilon})^{-1} \omega  )^{-1} P_+ | v_2 \rangle, 
%\label{AngularIntegralPerturbative}
%\end{eqnarray}

The integration over the angular rotations, taken to quadratic order, divides the functional integral by  two spectral determinants:
\begin{equation} \label{AngularIntegrationDeterminants}
\Pi_{\lambda_k \neq 0} (\lambda_k \pm \imath \omega (2 \pi \hat{\rho} \tilde{\epsilon} )^{-1} ) \times  \Pi_{\lambda_k \neq 0} (\lambda_k - \imath \omega (2 \pi \hat{\rho} \tilde{\epsilon} )^{-1} )
\end{equation}
Here $\lambda_k$ are the eigenvalues of the kinetic operator $\kappa$, and the spectral determinant expressed in $\Pi_{\lambda_k \neq 0}$ includes contributions from all of $\kappa$'s eigenvalues except the  zero eigenvalue associated with spatially uniform rotations, which is absent because we integrated global rotations of angular variables separately from their fluctuations.
%$\gamma_{UT} = (4 N \pi \hat{\rho}^2)^{2 - 2V} {\det}^{-1}(\kappa(1-\kappa)^{-1} P_+ \pm {\imath P_+ \omega  }  (2 \pi \hat{\rho} \tilde{\epsilon} )^{-1} ) \; {\det}^{-1}(\kappa  P_+ - \imath  P_+ \omega  (2 \pi \hat{\rho} \tilde{\epsilon} )^{-1}) $.

In addition to these spectral  determinants from the quadratic part of the Lagrangian, there is an important contribution from the zeroth order Lagrangian, i.e. the value of Lagrangian when the rotations are set to the identity and the  eigenvalues are fixed at their saddle point values.  This Lagrangian simplifies to
\begin{equation} \label{WignerDysonExponential}
e^{\imath \pi N V \hat{\rho} \omega \tilde{\epsilon}^{-1} (1 \pm 1)}
\end{equation}
where $\omega = E_1 - E_2 $ is the energy splitting and the $\pm$ sign denotes which saddle point is being used.  The Lagrangian also includes factors of $N$ and $e$ which are however cancelled out by factors from the functional integral's normalization and from integrations over $\overline{Q}^f$ and $\overline{Q}^b$'s  eigenvalues and rotations.  The exponential in Eq. \ref{WignerDysonExponential} encodes Wigner-Dyson oscillations, a well known manifestation of the repulsion between adjacent energy levels that occurs in disordered materials.

\subsection{The Determinant \label{FermionDeterminant}}

In the previous sections we have performed a saddle point analysis of the $\overline{Q}^f, \, \overline{Q}^b$ spins.  Fluctuations in the eigenvalues are of size $N^{-1/2}$ and are therefore suppressed in the both the $K$ fixed and $N\,K$ fixed limits as long as $N \rightarrow \infty$.  Fluctuations in the angles have size $(N \,K)^{-1/2}$ and therefore are suppressed in the $K$ fixed limit, deleting the localized phase.   In the $N\,K$ fixed limit angular fluctuations are not suppressed and the localized phase is preserved.  In addition,  global angular rotations have no mass except that caused by the energy splitting $\omega$.

In this section we analyze the  $ {\det({ \overline{Q}^f_{v_1}  {(1 - \kappa)^{-1/2}_{v_1 v_2}} {\delta_{j_1 j_2}} } - {  \overline{Q}^b_{v_1} L  (1 - \kappa)^{1/2}_{v_1 v_2} {\delta_{i_1 i_2}}  }   )}$ determinant which figures so prominently in the functional integral.    This determinant has the same mass structure as the  $\overline{Q}^f, \, \overline{Q}^b$  spins.  This can be seen by setting the angular fluctuations to unity, in which case the determinant factors into four determinants: 
\begin{eqnarray}
&\,& \det(x^f_{v_1,1} {(1 - \kappa)^{-1/2}_{v_1 v_2}} - x^b_{v_1,1}  (1 - \kappa)^{1/2}_{v_1 v_2} ) \nonumber \\
&\times& \det(x^f_{v_1,1} {(1 - \kappa)^{-1/2}_{v_1 v_2}} + x^b_{v_1,2}  (1 - \kappa)^{1/2}_{v_1 v_2} ) \nonumber \\
&\times & \det(x^f_{v_1,2} {(1 - \kappa)^{-1/2}_{v_1 v_2}} - x^b_{v_1,1} (1 - \kappa)^{1/2}_{v_1 v_2} ) \nonumber \\
&\times& \det(x^f_{v_1,2} {(1 - \kappa)^{-1/2}_{v_1 v_2}} + x^b_{v_1,2} (1 - \kappa)^{1/2}_{v_1 v_2} ) 
\end{eqnarray} 
The only differences between the four determinants are in which of the two $x^f$ eigenvalues contribute, and which of the two $x^b$ eigenvalues contribute.  

Recall that at the $+$ saddle point the $x^f$ values are the same as the $x^b$ ones i.e. $x^f_1 = x^b_1$ and $x^f_2 = x^b_2$, while at the $-$ saddle point they are interchanged i.e. $x^f_1 = x^b_2$ and $x^f_2 = x^b_1$.    Therefore in two of the determinants $x^f$ and $x^b$ nearly cancel each other and leave a term proportional to the energy splitting $\omega=E_1-E_2$, plus fluctuations in the saddle point which we have seen are of order $N^{-1/2}$, plus of course also a contribution from $\kappa$.  Therefore these two near-zero determinants are 
\begin{equation} \label{TopSpectralDeterminants}
{\det}^2(\kappa +  \imath (\pm 1 - 1) \omega ((2 \pi \hat{\rho} \tilde{\epsilon} )^{-1} ) /2 + \delta x^f - \delta x^b)
\end{equation}
where $\delta x^f, \; \delta x^b$ represent fluctuations in the saddle point, and the $\pm $ sign specifies which saddle point is in use.  These two determinants contain physics which is exactly the same as the non-uniform angular rotations of $\overline{Q}^f, \, \overline{Q}^b$.  

Since the kinetic operator's action on the constant vector is to multiply it by zero, each of these two determinants also contains a factor where the $\kappa$ disappears leaving only two factors of  $(\imath (\pm 1 - 1) \omega ((2 \pi \hat{\rho} \tilde{\epsilon} )^{-1} ) /2 + \delta x^f - \delta x^b)$.  The two factors contain exactly the same physics as global rotations of $\overline{Q}^f, \, \overline{Q}^b$.

Lastly there are two more determinants where $x^f$ and $x^b$ add to each other instead of nearly cancelling each other. The arguments of these determinants are roughly $\hat{\rho}+ \kappa$. The kinetic operator $\kappa$ is small compared to $\hat{\rho}$ which is of order one, so these two determinants represent purely local physics and are of order one.  In fact the two determinants represent the same physics seen in the eigenvalues of $\overline{Q}^f, \, \overline{Q}^b$.

\section{Summary}
In summary,  we have presented evidences that two disordered models, each with $N$ orbitals per site, possess a localized phase with finite localization length in the double scaling limit where $N \rightarrow \infty$ and $N\,K$ is kept fixed.  In the site-diagonal-disorder model we directly computed the localization length and showed that it remains finite in one dimension when $N\,K$ is kept fixed.  In the gauge invariant model, in any dimension or geometry, we showed that the $N\,K$ fixed limit preserves dynamic spin variables which in the $K$ fixed limit become very massive and are effectively deleted.  These spin variables may be capable of exiting the spontaneous symmetry breaking phase which occurs in their absence, and thus entering the Anderson localized phase.   These arguments indicate that $N$-orbital models are capable of exhibiting both localized and conducting phases, and an Anderson transition, in the double scaling  $N\,K$ fixed limit.

\begin{acknowledgments}
We especially thank Universit\`a degli Studi di Roma "La Sapienza", the Asia Pacific Center for Theoretical Physics, Sophia University, and  the Zhejiang Institute of Modern Physics, where much of this work was done.   We thank T. Ohtsuki for help during the early stages of this work.    We also acknowlege support from the Institute of Physics of the Chinese Academy of Sciences, Nanyang Technological University, Royal Holloway University of London, and Rutgers University.  We also thank  X. Wan, G. Parisi, T. Spencer, S. Kettemann, J. Zaanen, D. Culcer, P. Fulde, and A. Leggett for hospitality and discussion, and J. Verbaarschot, J. Keating, A. Mirlin, K. Efetov, M. Disertori, M. Zirnbauer, M. A. Skvortsov, K.-S. Kim, B. Liu, Y. Fyodorov, M. Milletari, C. Miniatura, B. Gremaud, and J. Y. Yong'En for helpful conversations.  
 % Also hosted by Les Houches, the Isaac Newton Institute, the Universit\`a degli Studi di Roma "La Sapienza", the ICTP,  the APCTP, and the IOP. 
%,  and acknowledges the Korea Ministry of Education, Science and Technology (MEST) for supporting the Young Scientist Training Program at the APCTP.  
% This work was supported by the National Science Foundation of China and by the 973 program of China  under Contract No. 2011CBA00108.
 \end{acknowledgments}

\begin{appendix}
\section{Calculation of the two point correlator}
% Many of the details here came from ReproSigmaModelCleanedUp.pdf, page 38.
Here we discuss how the hybrid functional integral can be used to easily calculate the two point density correlator.  First we collect leading order results mentioned in the text. The result of integrating the angular fluctuations of $\overline{Q}^f,\,\overline{Q}^b$ is division by two spectral determinants, given in Eq. \ref{AngularIntegrationDeterminants}:
\begin{equation} \label{AngularIntegrationDeterminantsCopy}
\Pi_{\lambda_k \neq 0} (\lambda_k \pm \imath \omega (2 \pi \hat{\rho} \tilde{\epsilon} )^{-1} ) \times  \Pi_{\lambda_k \neq 0} (\lambda_k - \imath \omega (2 \pi \hat{\rho} \tilde{\epsilon} )^{-1} )
\end{equation}
Here $\lambda_k$ are the eigenvalues of the kinetic operator $\kappa$, and the spectral determinant expressed in $\Pi_{\lambda_k \neq 0}$ includes contributions from all of $\kappa$'s eigenvalues except the zero eigenvalue associated with spatially uniform rotations.  The $\pm$ sign denotes which of the two saddle points we are computing. 

The Harish-Chandra-Itzykson-Zuber integral of spatially uniform rotations of $\overline{Q}^f$, and its analogue for $\overline{Q}^b$'s rotations, contribute a factor of  $\pm \omega^{-2} (2 \pi \hat{\rho} \tilde{\epsilon} )^{2}$. 

Equation \ref{TopSpectralDeterminants} gives the sector of the $\overline{Q}^f - \overline{Q}^b$ determinant which corresponds to angular rotations:
\begin{equation} \label{TopSpectralDeterminantsCopy}
\Pi_{\lambda_k } (\lambda_k +  \imath (\pm 1 - 1) \omega (4 \pi \hat{\rho} \tilde{\epsilon} )^{-1} + \delta x^f - \delta x^b )^2 
\end{equation}
The spatially uniform eigenvalue of $\kappa$, $\lambda = 0$, is included in this determinant. 

The value of the Lagrangian at the saddle point is $e^{\imath \pi N V \hat{\rho} \omega \tilde{\epsilon}^{-1} (1 \pm 1)}$. 

Next we put these results together. The eigenvalue fluctuations $\delta x^f,\,\delta x^b$ can be safely dropped where $\kappa$'s eigenvalues $\lambda_k$ are not zero.  We need  retain only the global averages of the eigenvalue fluctuations, $\delta \overline{x}^f,\,\delta \overline{x}^b$,  which contribute to the terms where $\lambda_k=0$.  We obtain:
\begin{eqnarray}
{\bar{Z}} & = & \sum_{\pm}\;\pm  e^{\imath \pi N V \hat{\rho} \omega \tilde{\epsilon}^{-1} (1 \pm 1)}
\nonumber \\
& \times& \frac{ (\imath \exp^{\imath \phi} (\pm 1 - 1) \omega (4 \pi \hat{\rho} \tilde{\epsilon} )^{-1} + \delta \overline{x}^f_1 - \delta \overline{x}^b_1 ) }{\omega (2 \pi \hat{\rho} \tilde{\epsilon} )^{-1}}
 \nonumber \\
&\times &  \frac{ (\imath \exp^{-\imath \phi} (\pm 1 - 1) \omega (4 \pi \hat{\rho} \tilde{\epsilon} )^{-1} + \delta \overline{x}^f_2 - \delta \overline{x}^b_2) }{\omega (2 \pi \hat{\rho} \tilde{\epsilon} )^{-1}}
\\ \nonumber 
&\times&  \Pi_{\lambda_k \neq 0}  \frac{ (\lambda_k +  \imath (\pm 1 - 1) \omega (4 \pi \hat{\rho} \tilde{\epsilon} )^{-1}  )^2 }    {  (\lambda_k \pm \imath \omega (2 \pi \hat{\rho} \tilde{\epsilon} )^{-1})  (\lambda_k - \imath \omega (2 \pi \hat{\rho} \tilde{\epsilon} )^{-1} )}
\end{eqnarray}
The phases $ \exp^{\pm \imath \phi}$ come from the saddle point values of the eigenvalues $ x =  \exp^{\pm \imath \phi}$, where $\phi$ is chosen so that $ \exp^{\pm \imath \phi} =  \sqrt{1 -(E/2\tilde{\epsilon})^2} \pm \imath E/ 2 \tilde{\epsilon}$.  The factors of $\overline{Q}^f$ and $\overline{Q}^b$ inside the determinant produce these phase factors.

 The $+$ and $-$ saddle points simplify to:
\begin{eqnarray}
{\bar{Z}}_- & = & -\, \frac{ (\imath \exp^{\imath \phi} \omega (2 \pi \hat{\rho} \tilde{\epsilon} )^{-1} + \delta \overline{x}^f_1 - \delta \overline{x}^b_1 ) }{\omega (2 \pi \hat{\rho} \tilde{\epsilon} )^{-1}}
 \nonumber \\
&\times &  \frac{ (\imath \exp^{-\imath \phi} \omega (2 \pi \hat{\rho} \tilde{\epsilon} )^{-1} + \delta \overline{x}^f_2 - \delta \overline{x}^b_2) }{\omega (2 \pi \hat{\rho} \tilde{\epsilon} )^{-1}}
\nonumber \\
{\bar{Z}}_+ & = &  e^{\imath  2 \pi N V \hat{\rho} \omega \tilde{\epsilon}^{-1} }
\nonumber \\
& \times& \frac{ (\delta \overline{x}^f_1 - \delta \overline{x}^b_1 ) ( \delta \overline{x}^f_2 - \delta \overline{x}^b_2) }{\omega^2 (2 \pi \hat{\rho} \tilde{\epsilon} )^{-2}}
\\ \nonumber 
&\times& \Pi_{\lambda_k \neq 0}  \frac{  \lambda_k^2 }    { (\lambda_k + \imath \omega (2 \pi \hat{\rho} \tilde{\epsilon} )^{-1})  (\lambda_k - \imath \omega (2 \pi \hat{\rho} \tilde{\epsilon} )^{-1} )}
\end{eqnarray}
This has the desired property that ${\bar{Z}}$'s average value is one, because single instances of the eigenvalue fluctuations average to zero, zero-ing the $+$ saddle point and setting the $-$ saddle point to one.  Because of the functional integral's construction, it should be identical to one for all values of its parameters, as long as factors of $\overline{Q}^b$ are not inserted to obtain density averages.

To calculate the density at point $v$, we must calculate the average of  
\begin{equation} 
-\imath N \tilde{\epsilon}^{-1} Q^b_{v,2} L =-\imath N \tilde{\epsilon}^{-1} (x^b_2 + \tilde{x}^b_{2,v} - 2 \pi \hat{\rho}((y^b_v)^2 + (z^b_v)^2)),
\end{equation} 
and take the real part.  $ \overline{Q}^b_{v,2} L $ is the lower right entry in $\overline{Q}^b_v L $, which is a $2 \times 2$ spin located at site $v$.  $x^b_2=-\exp^{-\imath \phi}$ is the saddle point value of $\overline{Q}^b$'s second eigenvalue multiplied by $L= \sigma_z$'s second eigenvalue, and $ \tilde{x}^b_{2,v}$ is the fluctuation in the second eigenvalue at position $v$.  Even though this fluctuation pairs with an instance of $ \delta \overline{x}^b_2$ in ${\bar{Z}}_+$, there is no pair for the $ \delta \overline{x}^b_1$, so contributions from eigenvalue fluctuations to the density at point $v$ are zero at leading order, and the $+$ saddle point contributes nothing.  At leading order the density at point $v$ is equal to  $N \tilde{\epsilon}^{-1} Re(\exp^{-\imath \phi}) = N \tilde{\epsilon}^{-1} \hat{\rho}$.   The  angular fluctuations $((y^b_v)^2 + (z^b_v)^2)$ dress this leading order result with a term proportional to the on-site value of the Green's function, i.e. ${Tr}((2 \pi \hat{\rho} \tilde{\epsilon} \kappa - \imath \omega)^{-1})$.  This represents a diffuson returning to its origin and dressing the density. 

Going on to the two point correlator, we now have to calculate the average of the product of  $-\imath N \tilde{\epsilon}^{-1} \overline{Q}^b_{v_1,2} L $ (discussed above) and $-\imath N \tilde{\epsilon}^{-1} \overline{Q}^b_{v_2,1} L$.  The saddle point value of this last factor's eigenvalue is $+ \exp^{\imath \phi}$.  There are three interesting terms in this product, two of which come from the $-$ saddle point, and the other from the $+$ saddle point.  First, the saddle point values $-\imath N \tilde{\epsilon}^{-1} x^b_{2,v}$ and $-\imath N \tilde{\epsilon}^{-1}  x^b_{1,v}$ multiply to give $N^2 \tilde{\epsilon}^{-2}$, which is multiplied by ${\bar{Z}}$'s average value, one.  Second,  the angular fluctuations at site $v_1$ correlate with the angular fluctuations at site $v_2$, producing a factor $((y^b_1)^2 + (z^b_v)^2) \times ((y^b_v)^2 + (z^b_v)^2)$.  After performing the integral, this turns into a product of two Green's functions connecting $v_1$ to $v_2$, i.e. $|\langle v_1 | (2 \pi \hat{\rho} \tilde{\epsilon} \kappa - \imath \omega)^{-1} | v_2 \rangle |^2$.  If one averages over position $v_1,\,v_2$ to get the correlator between energy levels $E_1,E_2$, this turns into ${Re} \,\,   {Tr}     ( (  2 \pi \hat{\rho}\, \tilde{\epsilon} \kappa - \imath  \omega  )^{-2}) $.  Both of these terms involve only the $-$ saddle point, not the $+$ saddle point. 

The two point correlator does have a contribution from the $+$ saddle point: the eigenvalue fluctuations $\tilde{x}^b_{1,v}$ and $\tilde{x}^b_{2,v}$ both pair with the $ \delta \overline{x}^b_1$ and  $ \delta \overline{x}^b_2$ in that saddle point, contributing a factor of $N^{-2}$.    Therefore the $+$ saddle point contributes to the two point correlator a term proportional to 
\begin{eqnarray}
& \;&  Re [ \omega^{-2} e^{\imath  2 \pi N V \hat{\rho} \omega \tilde{\epsilon}^{-1} }
\\ \nonumber 
&\times& \Pi_{\lambda_k \neq 0}  \frac{  \lambda_k^2 }    { (\lambda_k + \imath \omega (2 \pi \hat{\rho} \tilde{\epsilon} )^{-1})  (\lambda_k - \imath \omega (2 \pi \hat{\rho} \tilde{\epsilon} )^{-1} )} ]
\end{eqnarray}

With a little more care, and the choice of kinetic operator $\kappa = (-2 \pi \hat{\rho})^{-1} \nabla^2$, it can be shown that these three terms reproduce the standard result for the two level correlator $R_2$, which was obtained by Andreev, Altshuler, and Shklovskii \cite{Altshuler86, Andreev95, Andreev96}.  In our notation, their result is: 
\begin{eqnarray}
R_{2}(\omega)  - 1& = & (2 \pi^2 \rho^2)^{-1}  {Re} \,\,   {Tr}     ( (  2 \pi \hat{\rho}\, \tilde{\epsilon} \kappa - \imath  \omega  )^{-2}) 
\nonumber \\
&+& (2 \pi^2 \rho^2)^{-1}  \omega^{-2}    \cos(2  \pi N V \hat{\rho} \, \omega) 
\nonumber \\
& \times & \Pi_{\lambda_k \neq 0}  \frac{  \lambda_k^2 }    { (\lambda_k + \imath \omega (2 \pi \hat{\rho} \tilde{\epsilon} )^{-1})  (\lambda_k - \imath \omega (2 \pi \hat{\rho} \tilde{\epsilon} )^{-1} )} ,
\nonumber \\
  \pi \hat{\rho} \kappa &=& -\frac{D}{2} \nabla^2
\label{AndreevAltshulerCorrelator1}
\end{eqnarray}
This reduces to the standard Wigner-Dyson result in $D=0$ dimensions.   \footnote{We have shown calculation of the advanced-retarded correlator.  In order to obtain $R_2$ it is necessary to calculate also the much simpler advanced-advanced correlator, which contributes a factor of $-N^2 \tilde{\epsilon}^{-2} e^{-2 \imath \bar{\phi}}$.}

\section{The Jacobian for Global Rotations \label{JacobianAppendix}}

When the energy splitting $\omega = E_1 - E_2$ is equal to zero, the functional integral under study (see Eqs. \ref {DisertoriModel1}, \ref {SymmetrizedModel}) is exactly symmetric under global rotations where  all matrix spins $\overline{Q}^f,\,\overline{Q}^b$ rotate together.  The only terms which break this symmetry are the matrix elements of $\overline{Q}^b$ which one  inserts in order to measure observables, and also the energy splitting terms $ {{\imath N}{\tilde{\epsilon}}^{-1} \sum_v {Tr}(\overline{Q}^b_v L  {\hat{E}}   )} $ and $ {{\imath N}{\tilde{\epsilon}}^{-1} \sum_v {Tr}(  \overline{Q}^f_v  {{\hat{E}}} )}$.  This global continuous symmetry expresses mathematically the fact that two eigenstates of the disordered hamiltonian which are at the same energy necessarily are in fact one and the same eigenstate, and therefore are perfectly correlated with each other.  This symmetry is responsible for the existence of massless variables in the theory - the angular variables - which mediate electronic conduction. 

In the conducting phase all angular variables at all sites in the lattice are locked to each other, so  differences between their values at any two sites are small and controlled by powers of $(N\,K)^{-1}$.  Therefore we use a perturbative technique to integrate the spatial fluctuations of the angular variables.  

In contrast, global rotations of the angular variables are not small unless the level splitting $\omega$ is large compared to the level spacing $\Delta E$, i.e. $\omega \gg \Delta E$. Therefore global rotations must be integrated in a non-perturbative manner, exploring the complete manifold.  In consequence, it is necessary to make a change of variables which differentiates global rotations from fluctuations in the angular variables.   With the change of variables it is necessary to introduce a Jacobian, which is our focus here.

There are actually two changes of variables, one for the $\overline{Q}^f$ spins, and another for the $\overline{Q}^b$ spins.  The math for the former is very similar to the math for the latter, but in the end the Jacobian for $\overline{Q}^f$'s change of variables exhibits a divergence which is problematic, while $\overline{Q}^b$'s Jacobian is well behaved.  We will focus here on $\overline{Q}^f$ and its Jacobian.

We begin our analysis of the functional integral by decomposing the matrix spins  $\overline{Q}^f,\,\overline{Q}^b$ into eigenvalues and rotations.  The decomposition of  $\overline{Q}^f$ is $\overline{Q}^f_v = U_v x^f_v U_v^\dagger$, where the eigenvalue matrix $x^f_v$ is  diagonal, the rotation $U_v$ is unitary, and $v$ is the position index.  The decomposition of $\overline{Q}^bL$ is $\overline{Q}^b_v L = T_v x^b_v T^{-1}_v$, where $x^b$ is diagonal and constrained by $x^b L \geq 0$,  and $T$ is a member of the pseudo-unitary hyperbolic group $U(1, 1)$ defined by $T^\dagger L T = L$.     \cite{Fyodorov02}  $T$ lives on the  $U(1,1)$ manifold which is not compact and is indeed hyperbolic. 

We separate the global rotations from angular fluctuations using the variables $ U_v = U_0 \hat{U}_v, \;  T_v  = T_0 \hat{T}_v $, where $U_0,\, T_0$ are the global rotations and $\hat{U}_v,\,\hat{T}_v$ are the fluctuations.  We parameterize the fluctuations use the formula of Eq. \ref{PerturbativeParameterization}; the  parameterization of $\hat{U}_v$ is:
\begin{eqnarray}
\;\;\hat{U}_v &=&   \begin{bmatrix} {\sqrt{1 - (y^f_v)^2 - (z^f_v)^2}} & {\imath y^f_v + z^f_v}  \\ {\imath y^f_v - z^f_v}  & {\sqrt{1 - (y^f_v)^2 - (z^f_v)^2}} \end{bmatrix}
%\label{PerturbativeParameterization}
\nonumber
\end{eqnarray}
$y^f_v$ and $z^f_v$ vary within the unit circle defined by $(y^f_v)^2 + (z^f_v)^2 = 1$, and the  integration measure is $ {dy^f_v} {dz^f_v} $.
It is necessary add a constraint prohibiting global rotations in the fluctuation variables $\hat{U}_v$; we adopt  the requirement that $\sum_v y^f_v = 0$ and that $\sum_v z^f_v = 0$. 

We use the same parameterization, minus the global constraint, for the original angular variables $U_v,\;T_v$.  

We parameterize the global rotations using
\begin{eqnarray}
U_0 &=&  \begin{bmatrix} \cos \psi^f & \imath \sin \psi^f e^{-\imath \theta^f}  \\ \imath \sin \psi^f e^{\imath \theta^f}  & \cos \psi^f \end{bmatrix}, \nonumber \\
T_0 &=& \begin{bmatrix} \cosh \psi^b & \sinh \psi^b e^{-\imath \theta^b} \\ \sinh \psi^b e^{\imath \theta^b} & \cosh \psi^b  \end{bmatrix}
\end{eqnarray}
$\psi^b$ varies from $1$ to $\infty$,  $\psi^f$ varies from $0$ to $\pi /2$, and $\theta^f,\, \theta^b$ vary from $0$ to $2 \pi$.    The  Jacobian is $\frac{1}{4}\sin(2 \psi^f) \sinh (2 \psi^b)$, and this must be multiplied by another $1/2$ because the $Q^f$ parameterization covers its domain twice.

 We formally group the variables $\left[ \psi^f, \theta^f \right]$ of the global rotations into the vector $u_0$, the variables $\left[ y^f_v, z^f_v \right]$ of the original rotations into $u_v$, and the variables $\left[ y^f_v, z^f_v \right]$ of the angular fluctuations into $\hat{u}_v$.  The global constraint is $\sum_v \hat{u}_v = 0$.  The Jacobian of the change of variables is $J = \det(j_{v_1 v_2})$, where $j_{v_1 v_2}$ is the Jacobian matrix of derivatives.  This matrix includes a square component $\hat{j}_{v_1 v_2} = \frac{d u_{v_1}}{d \hat{u}_{v_2}}$ where the original variables $u_{v_1}$ are differentiated with respect to the angular fluctuation variables $\hat{u}_{v_2}$.  Only the diagonal components $\hat{j}_{v v}$ are non-zero.  In addition the Jacobian matrix is augmented with an extra column $\overline{j}_{v_1} = \frac{d u_{v_1}} {d u_0}$ where the original variables are differentiated with respect to the global rotation variables.
% Simple analysis of our constraint  $\sum_v \hat{u}_v = 0$ shows that the matrix inside the Jacobian is $B_U = \hat{j} + ( \overline{j} - \hat{j} | 0 \rangle )\otimes \langle 0 |$, where  $| 0 \rangle $ is the spatially uniform vector that satisfies $\langle v_1 | 0 \rangle = V^{-1/2}$.   Using the fact that   for every value of $U_0$ $\det(\hat{j}) = 1$, the Jacobian simplifies to $J = V^{-1} \det(B_U) , \; B_U =  \sum_{v} \hat{j}_{vv}^{-1} \overline{j}_v $. 
% at an intermediate step $B_U = 1 + (\hat{j}^{-1} \overline{j} -  | 0 \rangle )\otimes \langle 0 |$.  
%This simple form  implies that $J=1$ when the fluctuations $\hat{u} $ are zero.  The sum $\sum_v$ indicates that $J$ contains no contributions proportional to $\hat{u}$, and that the leading corrections are quadratic in $\hat{u}$; $J = 1 + O(\hat{u}^2)$.
Exploiting the structure of the constraint rapidly gives a simplified form for the Jacobian $J = V^{-1} \det(\sum_{v} \hat{j}_{vv}^{-1} \overline{j}_v )$.  The matrix inside this determinant no longer has spatial structure - all the spatial dependence has been subsumed into the sum over $v$, as a reward for having used a parameterization which afforded the very simple global restraint $\sum_v \hat{u}_v = 0$.

Expanding the Jacobian to quadratic  order in $\hat{y}_v, \hat{z}_v$, in preparation for   perturbation theory calculations, we obtain:
\begin{eqnarray}
\label{JacobianExpanded}
J & \approx & 1 - V^{-1} \sum_v 2({y}^f_v \hat{y}^f_v + {z}^f_v \hat{z}^f_v)  \nonumber \\
&+& V^{-1} \sum_v  ({y}^f_v \sin \phi^f - {z}^f_v \cos \phi^f)^2  a^f  \nonumber \\ 
& + &V^{-1} \sum_v 2({y}^b_v {y}^b_v + {z}^b_v {z}^b_v)  \nonumber \\
&-& V^{-1} \sum_v ({y}^b_v \sin \phi^b - {z}^b_v \cos \phi^b)^2  a^b , \\ \nonumber 
a^f &=& \frac{\sin^4 \psi^f (2 - \sin^2 \psi^f)}{(1 - \sin^2 \psi^f)^3}, \; a^b = \frac{ (1 + 2  \sin^{-2} \psi^b)}{(1 + \sin^{-2} \psi^b)^{3}}
%, \nonumber \\ r_0^2 &=& y_0^2 + z_0^2 = \sin^2 \psi ,  \; \sin \phi = z_0 / r_0
\end{eqnarray}

%The $a^f(r^f_0)$ function has a pole at $r^f_0 = 1$, signalling that our perturbative expansion of the Jacobian fails here, which is a problem because our integration over the saddle point manifold $U_0, \, T_0$ necessarily includes $r^f_0 = 1$.

The perturbative terms in the Jacobian are all weighted by a factor of the inverse system volume $V^{-1}$, making them quite small compared to other diagrams.  However they diverge when $  \psi^f = \pi/2$, which is the edge of the integration domain for the global rotations of $\overline{Q}^f$.   Intuitively the reason for this divergence is that angular rotations of $\overline{Q}^f$ are  limited to $\psi^f \leq \pi/2$, so when $\psi^f = \pi/2$ the angular fluctuations encounter a hard constraint allowing them to move from $\pi/2$ to smaller values, but not allowing them to move  to larger values.  The incompatibility between this constraint and the perturbative formulation of the angular fluctuations results in the Jacobian's divergence. This is the nonperturbative problem with the Jacobian which was  referred to in the main text of this article.  

As it stands, this form of the Jacobian is not useable and must be neglected, because if it is retained then the integration over the global rotations of $\overline{Q}^f$ will produce infinities resulting from the divergence of $a^f$.  The only  way to go forward while retaining the Jacobian is to require that the energy splitting $\omega$ be small compared to the level spacing $\Delta E$, in which case the global rotations will not explore the complete manifold of angular rotations.  This restriction would limit the work's usefulness for studying the metallic phase.

  The existence of the perimeter limiting the domain of integration, and the resulting divergence of the Jacobian, are caused by our parameterizations of the group $U(2)/U(1)\cdot U(1)$.  In the $\theta^f,\psi^f$ parameterization the $\psi^f$ variable is bounded to $\psi^f \leq \pi /2$, so that $\psi^f=\pi/2$ acts as a perimeter of the manifold.  A similar perimeter  exists at $(y^f)^2 + (z^f)^2 = 1$ in the $y^f,\; z^f$ parameterization.  However in both cases the perimeter is an artifact of the parameterization, and has no special meaning in the underlying group.  This is proved by the fact that it is  possible to adopt an alternative parameterization which is in some sense dual: the  perimeter in the original parameterization is the same as the center in the alternative parameterization, and vice versa. This transformation can be effected by setting $\sin \psi^f = \cos \hat{\psi}^f$.   In the alternative parameterization the Jacobian $\overline{Q}^f$ again diverges, but at the new perimeter, which corresponds to the center of the other parameterization.  The poles in $a^f$ can be avoided by adopting one parameterization for one part of the manifold and the alternative for the remainder of the manifold, but then $a^f$'s value is discontinuous at the boundaries between the two parameterizations, which casts doubt on the validity of this procedure.
  
  One promising way out of this quandary is to not enforce a global constraint on the collective behavior of rotations at all sites, and to instead constrain the angular variables at a particular site $v_0$, i.e. we could set $U(v = v_0) = T(v = v_0) = 1$.  I must attribute this approach to a suggestion from K. Efetov, but I have not been able to thoroughly explore its ramifications.  It is very attractive because the Jacobian is exactly equal to the identity so there is no possibility that it could cause problems in perturbation theory or elsewhere.  It also has a physical reasonableness.    The complication is that this approach explicitly breaks translational invariance, so that angular variables are absent at $v_0$.   This simple detail causes a significant change in the kinetic operator $\kappa$ which governs the angular fluctuations, because $\kappa$ is not allowed to mediate hopping to and from $v_0$.   Therefore a modified $\kappa$, minus $v_0$, would be used in  all formulas and terms in this paper which concern angular rotations.  The original kinetic operator would be retained in the determinant and when treating eigenvalues.
  
Other alternatives are more technical and less physical: exploration of other possibilities for parameterizing   $U(2)/U(1)\cdot U(1)$, exploration of other global constraints besides $\sum_v \hat{u}_v = 0$, or avoidance of the perturbative expansion of the Jacobian.  It may also be the case that in fully supersymmetric functional integrals the Jacobian is better behaved, or that the mathematics for properly handling any analogous divergence may have already been developed.  In this connection it is suspicious that in the approach adopted here there is a non-trivial Jacobian for angular rotations of $\overline{Q}^f, \overline{Q}^b$, but not for angular rotations of the Grassmann variables which are responsible for producing the determinant.

\end{appendix}

\bibliography{vincent}

\end{document}